\documentclass[sigconf]{acmart}

\usepackage{booktabs} % For formal tables
\usepackage{multirow}
\usepackage{color}
\usepackage{subfigure} % Written by Steven Douglas Cochran
\usepackage{array}
\usepackage{breqn}
\usepackage{{inputenc}}
\usepackage{balance}
\usepackage{multirow,tabularx}
\usepackage{balance}
\usepackage{longtable}
\usepackage{booktabs,longtable}
\usepackage{hhline}
\usepackage[flushleft]{threeparttable}
\usepackage{algorithm}
\usepackage{algorithmic}
\usepackage{epsfig}
\usepackage{graphicx}
\usepackage{epstopdf}
\usepackage{thmtools}
\usepackage{enumitem}
\usepackage{verbatim}
\usepackage{amsfonts}
\usepackage{hyperref}
\hypersetup{colorlinks=false,linkcolor=blue,urlcolor=blue,citecolor=red}
\newcommand\norm[1]{\left\lVert#1\right\rVert}
\newcommand{\Lapl}{\mathbf{\mathop{\mathcal{L}}}}
\newcommand{\Trans}[1]{{#1}^{\top}}

\newcommand{\Mat}[1]{\mathbf{#1}}

\newcommand{\Set}[1]{\mathcal{#1}}

\newcommand{\ie}{\emph{i.e., }}
\newcommand{\eg}{\emph{e.g., }}
\newcommand{\etal}{\emph{et al.}}

\newcommand{\wrt}{\emph{w.r.t. }}
\newcommand{\cf}{\emph{cf. }}
\newcommand{\aka}{\emph{aka. }}

\copyrightyear{2017}
\acmYear{2017}
\setcopyright{acmcopyright}
\acmConference{SIGIR’17}{}{August 7--11, 2017, Shinjuku, Tokyo, Japan}
\acmPrice{15.00}
\acmDOI{http://dx.doi.org/10.1145/3077136.3080771}
\acmISBN{978-1-4503-5022-8/17/08}

\begin{document}
\title{Item Silk Road: Recommending Items from Information Domains to Social Users}

\author{Xiang Wang}
\affiliation{%
  \institution{National University of Singapore}
  %\streetaddress{National University of Singapore}
  %\state{Singapore}
  %\postcode{117417}
}
\email{xiangwang@u.nus.edu}

\author{Xiangnan He}
\authornote{Xiangnan He is the corresponding author.}
\affiliation{%
  \institution{National University of Singapore}
}
\email{xiangnanhe@gmail.com}

\author{Liqiang Nie}
\affiliation{%
  \institution{ShanDong University}
}
\email{nieliqiang@gmail.com}

\author{Tat-Seng Chua}
\affiliation{%
  \institution{National University of Singapore}
}
\email{dcscts@nus.edu.sg}
%\renewcommand{\shortauthors}{Xiang et. al.}

\begin{abstract}
Online platforms can be divided into information-oriented and social-oriented domains. The former refers to forums or E-commerce sites that emphasize user-item interactions, like Trip.com and Amazon; whereas the latter refers to social networking services (SNSs) that have rich user-user connections, such as Facebook and Twitter.
Despite their heterogeneity, these two domains can be bridged by a few overlapping users, dubbed as \emph{bridge users}.
In this work, we address the problem of \textit{cross-domain social recommendation}, \emph{i.e.}, recommending relevant items of information domains to potential users of social networks.
To our knowledge, this is a new problem that has rarely been studied before.

Existing cross-domain recommender systems are unsuitable for this task since they have either focused on homogeneous information domains or assumed that users are fully overlapped. Towards this end, we present a novel \emph{Neural Social Collaborative Ranking} (NSCR) approach, which seamlessly sews up the user-item interactions in information domains and user-user connections in SNSs.
In the information domain part, the attributes of users and items are leveraged to strengthen the embedding learning of users and items. In the SNS part, the embeddings of bridge users are propagated to learn the embeddings of other non-bridge users. Extensive experiments on two real-world datasets demonstrate the effectiveness and rationality of our NSCR method.
%well-validate our scheme.
\end{abstract} \vspace{-5pt}


\vspace{-5pt}

 \vspace{-5pt}
\maketitle

\section{Introduction}
Nowadays online platforms play a pivotal role in our daily life and encourage people to share experiences, exchange thoughts, and enjoy online services.
%People treat various platforms as different information channels to fulfill their diverse needs.
Regardless of applications, we can roughly divide the existing platforms into information-oriented and social-oriented domains.
The former typically refers to forums or E-Commerce sites that have thorough knowledge on items, such as point-of-interests in Trip.com, movies in IMDb, and products in Amazon. These sites have ample user-item interactions available in the form of users' reviews, ratings, along with various kinds of implicit feedback like views and clicks~\cite{iCD}.
On the other hand, the social-oriented domains are mainly social network sites, which emphasize the social connections among users~\cite{SNE}.

When adopting an item, besides consulting the information sites, a user usually gathers more detailed information from her experienced friends. This refers to \emph{word-of-mouth marketing}, which is widely recognized as the most effective strategy for producing recommendation.
As reported by Cognizant\footnote{\url{https://www.cognizant.com}.}, more than $45\%$ of travelers rely on social networks to seek advice from friends for travel.
However, most existing SNSs, like Facebook and Twitter, are designed mainly for users to rebuild their real-world connections, rather than for seeking options regarding items.
Though some item cues implying users' preference can be found in SNSs, they typically contain item names only with limited details.
The sparse and weak user-item interactions greatly hinder the ability of SNSs to offer item recommendation services.
%Therefore, we believe routing relevant items from information domains to the potential users of social domains is an essential yet challenging task.

Fortunately, some users may be simultaneously involved in both SNSs and information-domain sites, who can act as a bridge to propagate user-item interactions across domains.
For example, it is not unusual for a user to share her travel experiences in Trip.com; and if the user also holds a Facebook account, we can recommend her friends in Facebook with her liked items from Trip.com.
In social circles, these bridge users are like the \emph{silk road} to route relevant items from information domains to  (non-bridge) users of social networks.
As such, we formulate the task of \textit{cross-domain social recommendation}, which aims to recommend relevant items of information domains to the users of social domains.
Apparently, this task is related to the recently emerging topic --- cross-domain recommendation~\cite{DBLP:journals/tkde/JiangCCW0Y15}.
However, we argue that existing efforts have either focused on homogeneous domains (\textit{i.e.,} multiple sites of the information domain)~\cite{DBLP:conf/www/ElkahkySH15}, or unrealistically assumed that the users are fully overlapped~\cite{DBLP:journals/tkde/JiangCCW0Y15,wang2017unifying}. Our task to address is particularly challenging due to the following two practical considerations.
%As a consequence, they can hardly be applied to our task due to the following key challenges,

\begin{itemize}[leftmargin=*]
	%XN: mentioned already. no need to repeat.
  %\item Heterogeneous domains.
  %By involving different types of relations,
 % It is evident that the user-item interactions in information domains and user-user social relations in SNSs are intrinsically different from each other. For example, the ratings regarding specific items from users in Trip.com and friendship relations between users in Facebook are exclusive correlations between different types of entities.
  \item Insufficient bridge users. To gain a deep insight, we analyzed the overlapped users between Trip.com and Facebook/Twitter, finding that only $10.5\%$ of $8,196$ Facebook users and $6.9\%$ of $7,233$ Twitter users have public accounts in Trip.com. It is highly challenging to leverage history of such limited number of bridge users to provide quality recommendation for non-bridge users.% In a sense, the non-bridge users of SNSs can be viewed as cold-start users for the information domain, since they have no historical interaction on the items. It is highly challenging to leverage the limited histories of bridge users to provide quality recommendation for non-bridge users.
  %Making full use of the insufficient bridge users is an encouraging but challenging problem.
  \item Rich attributes. The users and items of an information domain are usually associated with rich attributes. For instance, Trip.com enables users to indicate their travel preference explicitly, and associates travel spots (\textit{i.e.,} items) with specific travel modes, among other information. However, little attention has been paid to leverage these attributes to boost the performance of cross-domain recommendation.
\end{itemize} \vspace{-5pt}

\noindent In this work, we propose a novel solution named \emph{Neural Social Collaborative Ranking} (NSCR) for the new task of cross-domain social recommendation. It is developed based on the recent advance of neural collaborative filtering (NCF)~\cite{heneural}, which is further extended to model the cross-domain social relations by combining with the graph regularization technique~\cite{DBLP:conf/cikm/HeCKC15}. We entail two key technical components of our NSCR as follows. \begin{itemize}[leftmargin=*]
%To address the challenges above, we leverage on the recent advances in deep learning~\cite{heneural} by presenting a generic solution named\ \emph{Neural Social Collaborative Ranking} (NSCR) for cross-domain social recommendation. Our proposed framework seamlessly integrates the user-item interactions of information domains with the user-user connections of social domains.
\item For the modelling of information domain, we build an attribute-aware recommender based on the NCF framework. To fully exploit the interactions among a user, an item, and their attributes, we enhance NCF by plugging a \textit{pairwise pooling} operation above the embedding vectors of user (item) ID and attributes. In contrast to the default average pooling used by NCF~\cite{heneural} and other recent neural recommenders~\cite{DBLP:conf/recsys/CovingtonAS16}, our use of pairwise pooling better captures feature interactions in the low level~\cite{he2017neural,DBLP:conf/uai/RendleFGS09}, greatly facilitating the following deep layers to learn higher-order interactions among users, items and attributes.
%Benefiting from the designed \emph{pairwise pooling} operation, our framework efficiently models the user-attribute, item-attribute, and attribute-attribute correlations. In the light of this, we can strengthen the representation learning of users and items by leveraging the rich attributes. Distinct from factorization methods that estimate a preference score as the inner product between user and item representations, we instead feed their element-wise product into a deep neural network. As such, NSCR captures the complex higher-order correlations among users, items, and attributes, allowing a full neural network treatment for collaborative ranking.
\item For the modelling of social domain, it is natural to guide the embedding learning of social users by using the embeddings of bridge users. As the embeddings of bridge users are optimized to predict user--item interactions (\eg ratings and purchases), propagating their embeddings to social users helps to bridge the heterogeneity gap between information domain and social domain. To implement such propagation effect, we employ the \emph{smoothness} constraint (\ie\emph{graph Laplacian}) on the social network, which enforces close friends to have similar embedding so as to reflect their similar preferences.
\end{itemize}

\noindent To sum up, the key contributions of this work are three-fold:
\begin{enumerate}[leftmargin=*]
  \item To our knowledge, we are the first to introduce the task of cross-domain social recommendation, which recommends relevant items of information domains to target users of social domains.
  \item We propose a novel solution that unifies the strengths of deep neural networks in modelling attributed user-item interactions and graph Laplacian in modelling user-user social relations.
  %We propose a bilinear pooling method, which considers all pairwise correlations related to user-item interactions and encodes the second-order interactions between user/item attributes in the embedding space. We empirically demonstrate its high expressiveness in recommendation tasks.
  \item We construct two real-world benchmark datasets for exploring the new task of cross-domain social recommendation and extensively evaluate our proposed solution.
  %comprising of user-item interactions in Trip.com and user-user connections in Facebook/Twitter. Thereinto, Trip.com is treated as the forum and the others are viewed as SNSs. We have successfully evaluated the framework with a practical study of travel recommendation.
\end{enumerate}

\section{Preliminary}
We first formulate the task of cross-domain social recommendation, and then shortly recapitulate the matrix factorization model, highlighting its limitations for addressing the task.
\subsection{Problem Formulation}
\begin{figure}
	\centering
	\includegraphics[width=0.46\textwidth]{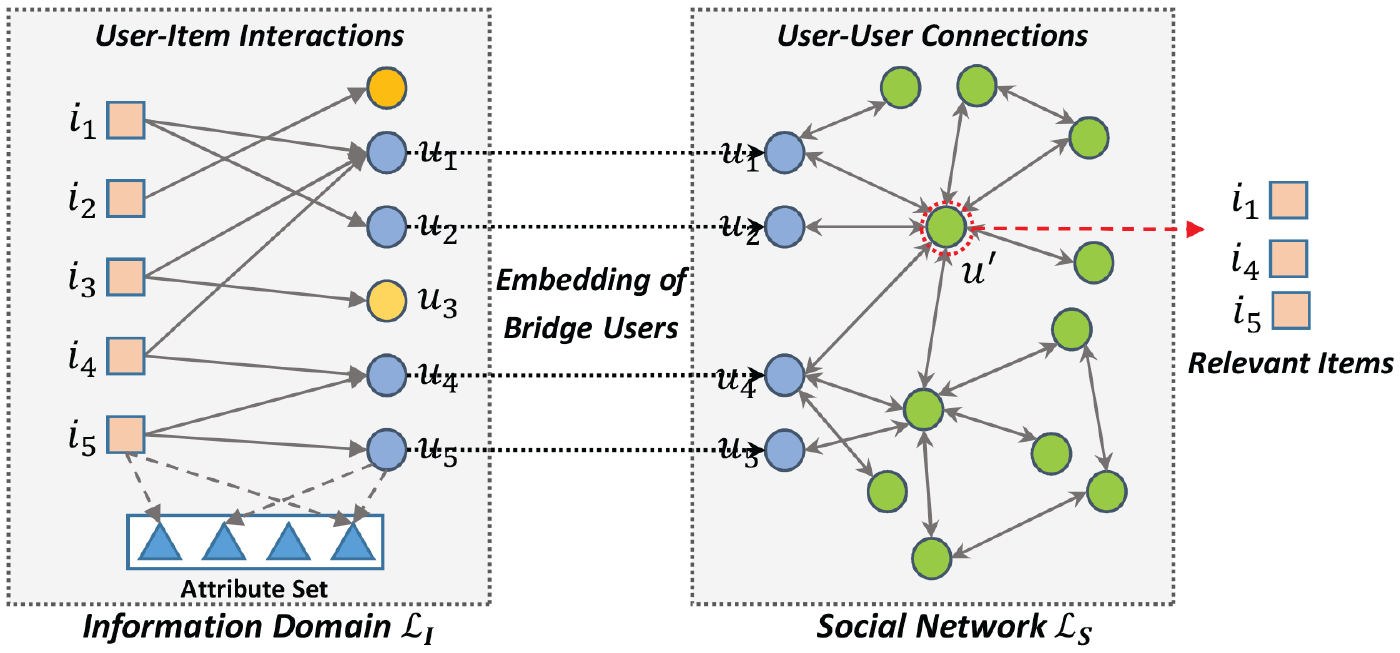}\\
	\vspace{-5pt}
	\caption{Illustration of the cross-domain social recommendation task.}\label{fig:framework}
	\vspace{-1em}
\end{figure}

Figure~\ref{fig:framework} illustrates the task of cross-domain social recommendation.
In the information domain, we have the interaction data between users and items.
Let $u$ and $\Set{U}_{1}=\{u_{t}\}_{t=1}^{M_{1}}$ denote a user and the whole user set of the information domain, respectively; similarly, we use $i$ and $\Set{I}=\{i_{t}\}_{t=1}^{N}$ to denote an item and the whole item set, respectively.
The edges between users and items denote their interactions, $\Set{Y} = \{y_{ui}\}$, which can be real-valued explicit ratings or binary 0/1 implicit feedback.
Traditional collaborative filtering algorithms can then be performed on the user-item interaction data.

In addition to the ID that distinguishes a user or an item, most information-domain sites also associate them with abundant side information, which can help to capture users' preferences and item properties better.
For example, in Trip.com, the user may choose the travel tastes of \emph{\{luxury travel, art lover\}} in her profile; while, the item \emph{Marina Bay Sands} is tagged most with travel modes \emph{\{luxury travel, family travel, nightlife\}}.
We term these associated information as \textit{attributes}, most of which are discrete categorical variables for the web domain~\cite{he2017neural}.
Formally, we denote $g$ and $\Set{G}=\{g_{t}\}_{t=1}^{V}$ as an attribute and the whole attribute set, respectively; for a user $u$ and an item $i$, we can then construct the associated attribute set as $\Set{G}_{u}=\{g^{u}_{1},\cdots,g^{u}_{V_{u}}\}\subset\Set{G}$ and $\Set{G}_{i}=\{g^{i}_{1},\cdots,g^{i}_{V_{i}}\}\subset\Set{G}$, respectively.

In the social domain, we have social connections between users, such as the undirected friendship or directed follower/followee relations.
We denote a social user as $u'$, all users of the social domain as $\Set{U}_{2}=\{u'_{t}\}_{t=1}^{M_{2}}$, and all social connections as $\Set{S}=\{s_{u' u''}\}$.
We define the bridge users as the overlapping users between the information domain and social domain. These bridge users can be expressed as $\Set{U}=\Set{U}_{1}\cap\Set{U}_{2}$.
In a social network, a user's behaviours and preferences can be propagated along the social connections to influence her friends. As such, these bridge users play a pivotal role in addressing the cross-domain social recommendation problem, which is formally defined as:
\begin{description}
	\item[\textbf{Input}:] An information domain with $\{\Set{U}_1, \Set{I}, \Set{Y}, \Set{G}_{u}, \Set{G}_{i}\}$; a social domain with $\{\Set{U}_2, \Set{S}\}$; and $\Set{U}_1\cap \Set{U}_2$ is nonempty.
	\item[\textbf{Output}:] A personalized ranking function for each user $u'$ of the social domain $f_{u'} : \Set{I} \to \mathbb{R}$, which maps each item of the information domain to a real number.
\end{description}
It is noted that there indeed exist sparse and weak user-item interactions in SNSs as aforementioned. However, we simplify this scenario of cross-domain social recommendation by only emphasizing the social connections in SNSs and leaving the exploration of weak interactions as the future work.

\subsection{Factorization Model}
Collaborative filtering (CF) is the key technique for personalized recommendation systems. It exploits user-item interactions by assuming that similar users would have similar preference on items.
%Depending on the way of measuring similarity, we can categorize CF approaches into two types: memory-based~\cite{DBLP:conf/www/SarwarKKR01} and model-based~\cite{DBLP:conf/kdd/Koren08}. While memory-based approaches tend to directly measure user/item similarity from the interaction data,
Model-based CF approaches~\cite{iCD,DCF} achieve this goal by describing the interaction data with an underlying model, for which the holistic goal is to build:
%, leading to more accurate prediction than memory-based approaches~\cite{DBLP:conf/www/SarwarKKR01}. Regardless of the underlying model, we can formulate the holistic goal of model-based approaches as:
\begin{equation}\label{equ:predicted-rating}
\widehat{y}_{ui}=f_{\Theta}(u,i),
\end{equation}
where %$u$ ($i$) denotes a user (an item),
$f$ denotes the underlying model with parameters $\Theta$, and $\widehat{y}_{ui}$ denotes the predicted score for a user-item interaction $y_{ui}$.
%SPACE. add back if you have more space or camera ready.
%Among the various models, such as graph-based~\cite{} and Bayesian networks~\cite{},
Matrix factorization (MF) is one of the simplest yet effective models for the recommendation task, which characterizes a user or an item with a latent vector, modelling a user-item interaction as the inner product of their latent vectors:
\begin{equation}
\label{equ:mf}
f_{MF}(u,i|\Mat{p}_{u},\Mat{q}_{i})=\Trans{\Mat{p}}_{u}\Mat{q}_{i}=\sum_{k=1}^{K}p_{uk}q_{ik},
\end{equation}
where $\textbf{p}_u\in\mathbb{R}^K$ and $\textbf{q}_i\in\mathbb{R}^K$ are model parameters denoting the latent vector (\aka representation) for user $u$ and item $i$, respectively.

\begin{figure}
	\centering
	\includegraphics[width=0.46\textwidth]{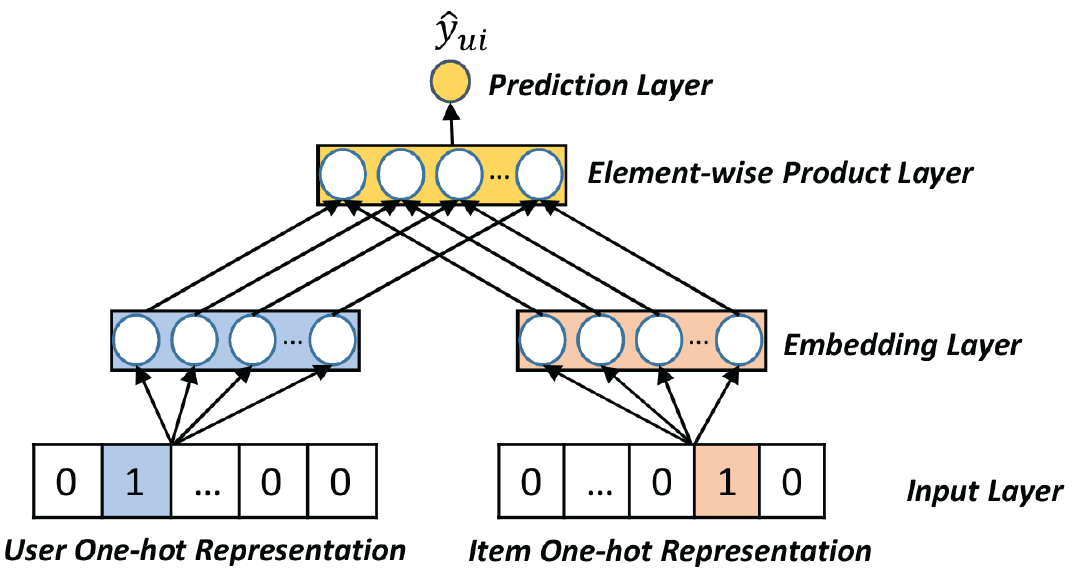}
	\vspace{-5pt}
	\caption{MF as a shallow neural network model.}\label{fig:mf_limit}
	\vspace{-1em}
\end{figure}

Despite its effectiveness, we note that MF's expressiveness can be limited by the use of the inner product operation to model a user-item interaction. To illustrate this, we present a neural network view of the MF model. As shown in Figure~\ref{fig:mf_limit},  we feed the one-hot representation of user/item ID into the architecture, and project them with a fully connected embedding layer. By feeding the user/item embedding vectors into the element-wise product layer, we obtain a hidden vector $\textbf{h}=\{p_{uk}q_{ik}\}$.
If we directly project $\textbf{h}$ into the output score, we can exactly recover the MF model.
As such, MF can be deemed as a shallow neural network with one hidden layer only. Based on this connection, we argue that there are two key limitations of MF-based approaches for cross-domain social recommendation:
\begin{itemize}[leftmargin=*]	
	\item First, MF only considers the simple two-way interaction between a user and an item, by assuming that their cross latent factors (\ie $\Mat{p}_{u}$ and $\Mat{q}_{i}$) are independent of each other. However, such an independence assumption can be insufficient to model real-world data,
	%XN: add the citation I used in my NFM paper.
	which usually have complex and non-linear underlying structures~\cite{he2017neural,SNE}.
	\item The case can be even worse if we take the attributes into account. A typical way to extend MF with side attributes is SVDfeature, \ie by summing attribute embedding vectors with user/item embedding vector. As a result, the rich correlations among users, items, and attributes are unintentionally ignored.
\end{itemize}

\noindent Our proposed NSCR solution addresses the above limitations of MF by 1) using a deep learning scheme to capture the higher-order correlations between user and item latent factors, and 2) devising a pairwise pooling operation to efficiently model the pair-wise correlations among users, items, and attributes.

\section{Our NSCR Solution}

The goal of cross-domain social recommendation is to select relevant items from the information domain for social users. Under the paradigm of embedding-based methods (\aka representation learning), the key for addressing the task is on how to project items (of the information domain) and users (of the social domain) into the same embedding space.
A generic solution is the factorization machine~(FM)~\cite{DBLP:conf/uai/RendleFGS09,DBLP:journals/tist/Rendle12}, which merges the data from the two domains by an early fusion; that is, constructing the predictive model by incorporating social users as the input features.
While the solution sounds reasonable conceptually, the problem is that the training instances which can incorporate social users are only applicable to the bridge users, which can be very few for real-world applications. As such, the generic recommender solution FM can suffer severely from the problem of insufficient bridge users.

To address the challenge of insufficient bridge users, we propose a new framework that separates the embedding learning process of each domain.
By enforcing the two learning processes to share the same embeddings for bridge users, we can ensure that items and social users are in the same embedding space.
Formally, we devise the optimization framework as:
\begin{gather}\label{equ:framework}
\Lapl = \Lapl_{I} (\Theta_I) + \Lapl_{S}(\Theta_S),
\end{gather}
where $\Lapl_{I}$ (or $\Lapl_{S}$) denotes the objective function of the information domain (or social domain) learning with parameters $\Theta_I$ (or $\Theta_S$), and most importantly, $\Theta_I \cap \Theta_S$ are nonempty denoting the shared embeddings of bridge users.

By separating the learning process for two domains, we allow the design of each component to be more flexible. Specially, we can apply any collaborative filtering solution for $\Lapl_I$ to learn from user-item interactions, and utilize any semi-supervised learning technique for $\Lapl_S$ to propagate the embeddings of bridge users to non-bridge users.
In the remainder of this section, we first present our novel neural collaborative ranking solution for $\Lapl_I$, followed by the design of social learning component $\Lapl_S$. Lastly, we discuss how to optimize the joint objective function.

%based on the predicted preference $\hat{y}_{u'i}$. Towards this end, we explore deep learning techniques and present a generic deep cross-domain social recommendation framework, which seamlessly integrate user-item interactions with accompanying attributes in information domains and user-user social connections in SNSs. The generic framework is formulated as,

%\begin{figure}
%  \centering
%  % Requires \usepackage{graphicx}
%  \includegraphics[width=0.48\textwidth]{Chart/overall.pdf}\\
%  \caption{Illustration of our generic framework.}\label{fig:framework}
%\end{figure}

%In the following sections, we first present a standard collaborative ranking (CR) model. Moving one more step towards our goal, we present a neural attribute-aware CR model, $\Lapl_{s}$, which unifies the attribute-aware user/item representation learning and the neural network structure in the information domain. We then propose the social network embedding, $\Lapl_{u}$, to learn representations for social users in SNSs. Lastly, we present how to recommend the item to social users.

\subsection{Learning of Information Domain}
\label{ss:learning_information}
To estimate the parameters for a CF model from user-item interaction data, two types of objective functions --- point-wise~\cite{iCD,heneural} and pair-wise~\cite{DBLP:conf/uai/RendleFGS09,chen2017acf,RankALS} --- are most commonly used.
The point-wise objective functions aim to minimize the loss between the predicted score and its target value.
%, which is mainly used for explicit feedback, such as user ratings. However, they are neither suited for learning from implicit feedback nor emphasizing the goal of personalized ranking.
%In many sites of information domains, implicit feedback are tracked automatically thus are more abundant, such as users' views on pages and purchases on products.
Here, to tailor our solution for both implicit feedback and the personalized ranking task, we adopt the pair-wise ranking objective functions.

Formally, we denote an observed user-item interaction as $y_{ui} = 1$, otherwise $y_{ui} = 0$. Instead of forcing the prediction score $\hat{y}_{ui}$ to be close to $y_{ui}$, ranking-ware objective functions concern the relative order between the pairs of observed and unobserved interactions:
\begin{gather}
\Lapl_I = \sum_{(u,i,j)\in\Set{O}}\Lapl(y_{uij},\hat{y}_{uij}),
\end{gather}
where $y_{uij}=y_{ui}-y_{uj}$ and $\hat{y}_{uij}=\hat{y}_{ui}-\hat{y}_{uj}$; $\Set{O}$ denotes the set of training triplets, each of which comprises of a user $u$, an item $i$ of observed interactions (\ie $y_{ui}=1$), and an item $j$ of unobserved interactions (\ie $y_{ui}=0$). An ideal model should rank all $(i,j)$ item pairs correctly for every user. To implement the ranking hypotheses, we adopt the regression-based loss~\cite{RankALS}:
\begin{equation}\label{equ:information-obj}
\Lapl_I=\sum_{(u,i,j)\in\Set{O}}(y_{uij}-\hat{y}_{uij})^{2} = \sum_{(u,i,j)\in\Set{O}} (\hat{y}_{ui} - \hat{y}_{uj} - 1)^2 .
\end{equation}
\noindent Note that other pair-wise ranking functions can also be applied, such as the bayesian personalized ranking~(BPR)~\cite{chen2017acf,DBLP:conf/uai/RendleFGS09} and contrastive max-margin loss~\cite{DBLP:conf/nips/SocherCMN13}. In this work, we use the regression-based ranking loss as a demonstration for our NSCR, and leave the exploration of other choices as the future work.

\subsubsection{\textbf{Attribute-aware Deep CF Model}}\label{sec:representation-learning}
\begin{figure}
	\centering
	\includegraphics[width=0.46\textwidth]{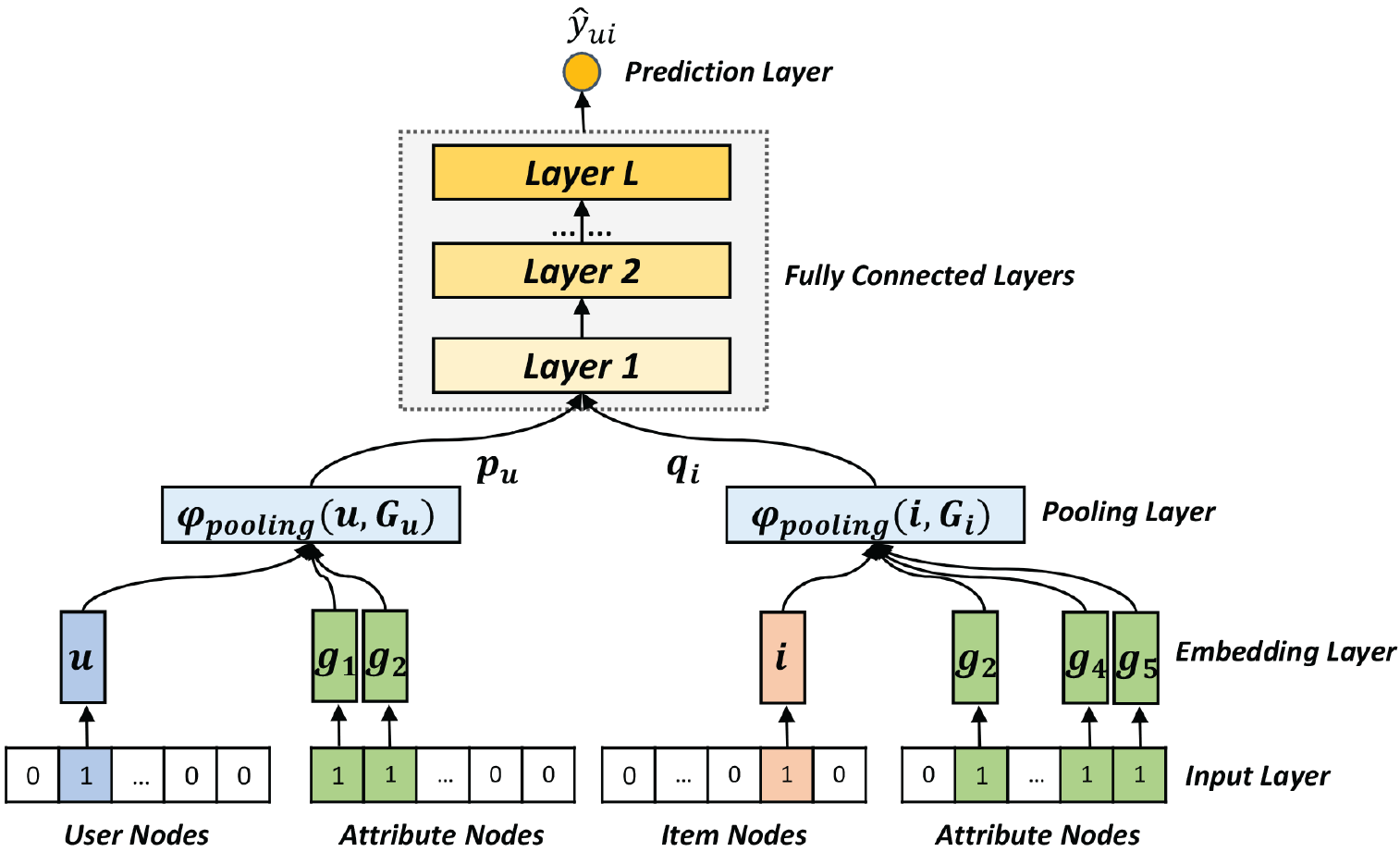}\\
	\vspace{-5pt}
	\caption{Illustration of our Attributed-aware Deep CF model for estimating an user-item interaction.}\label{fig:neural-collaborative-ranking}
	\vspace{-1em}
\end{figure}

Having established the optimization function for learning from information domain, we now present our attribute-aware deep collaborative filtering model to estimate a user-item interaction $\hat{y}_{ui}$. Figure~\ref{fig:neural-collaborative-ranking} illustrates its architecture, which is a multi-layered feed-forward neural network. We elaborate its design layer by layer.

%With the user/item representations at hand, we purpose to properly model the user-item interaction. To accomplish this task, we present a neural collaborative ranking (NCR) model to capture nonlinear and complex inherent structure of real-world interactions. We elaborate the design of NCR layer by layer, as Figure~\ref{fig:neural-collaborative-ranking} demonstrates.

\textbf{Input Layer.} The input to the model is a user $u$, an item $i$, and their associated attributes $\Set{G}_{u}$ and $\Set{G}_{i}$. We transform them into barbarized sparse vectors with one-hot encoding, where only the non-zero binary features are recorded.
%We use only the identities of a user and an item as the input feature, transforming them into binarized sparse vectors with one-hot-encoding, which is similar to the input feature as Figure~\ref{fig:mf_limit} demonstrates.

\textbf{Embedding Layer.}~The embedding layer maps each non-zero feature into a dense vector representation. As we have four types of features here, we differentiate them with different symbols: $\Mat{u}$, $\Mat{i}$, $\Mat{g}^u_t$, and $\Mat{g}^i_t$ denote the $K$-dimensional embedding vector for user $u$, item $i$, user attribute $g^u_t$, and item attribute $g^i_t$, respectively.

\textbf{Pooling Layer.}~The output of the embedding layer is a set of embedding vectors to describe user $u$ and item $i$, respectively. As different users (items) may have different number of attributes, the size of the embedding vector set may vary for different inputs. To train a neural network of fixed structure, it is essential to convert the set of variable-length vectors to a fixed-length vector, \ie the pooling operation.

The most commonly used pooling operations in neural network modelling are average pooling and max pooling. However, we argue that such simple operations are insufficient to capture the interaction between users/items and attributes.
For example, the average pooling assumes a user and her attributes are linearly independent, which fails to encode any correlation between them in the embedding space.
To tackle the problem, we consider to model the pairwise correlation between a user and her attributes, and all nested correlations among her attributes:
%To tackle the problem, inspired from our recent design of neural factorization machines~\cite{he2017neural}, we consider a similar pooling operation, named \textit{bilinear pooling}, to model all nested pair-wise correlation between a user and her attributes:

%users/items and attributes are linearly independent, which fails to encode any correlation between attributes and user/items.
%
%With the set of embedding vectors at hand, we feed them into our pooling layer, which can be set as the average pooling or bilinear pooling as discussed in Section~\ref{sec:representation-learning}. Moreover, the pooling layer enrich the user/item representations by considering their attributes. We incorporate the associated attributes into the user/item representations as,
%\begin{gather}
%\Mat{p}_{u}=\varphi(u,\Set{G}_{u}),\quad\Mat{q}_{i}=\varphi(i,\Set{G}_{i}),
%\end{gather}
%where $\varphi$ measures the correlation between a user (an item) and her (its) specific attributes. Here we entail one effective pooling strategy, converting a set of embedding vectors to one vector.
%
%\textbf{Pooling :}~Moving one more step, we model all nested interactions between the attributes and user embedding, dubbed as \emph{bilinear pooling}, as,
\begin{align}\label{equ:user-bilinear-pooling}
\Mat{p}_{u}=\varphi_{pairwise}(\Mat{u},\{\Mat{g}^u_t\})&=\sum_{t=1}^{V_{u}}\Mat{u}\odot\Mat{g}^{u}_{t}+\sum_{t=1}^{V_{u}}\sum_{t'=t+1}^{V_{u}}\Mat{g}^{u}_{t}\odot\Mat{g}^{u}_{t'},
\end{align}
where $\odot$ denotes the element-wise product of two vectors. We term it as \emph{pairwise pooling}, which is originally inspired from the design of factorization machines~\cite{DBLP:conf/icdm/Rendle10,he2017neural}. By applying pairwise pooling on the item counterpart, we can similarly model the pair-wise correlation between an item and its attributes:
\begin{align}\label{equ:item-bilinear-pooling}
\Mat{q}_{i}=\varphi_{pairwise}(\Mat{i},\{\Mat{g}^i_t\})&=\sum_{t=1}^{V_{i}}\Mat{i}\odot\Mat{g}^{i}_{t}+\sum_{t=1}^{V_{i}}\sum_{t'=t+1}^{V_{i}}\Mat{g}^{i}_{t}\odot\Mat{g}^{i}_{t'}.
\end{align}
%where this pooling method contains user-attribute, item-attribute, and attribute-attribute correlations. Clearly, the output of bilinear pooling is a $K$-dimensional vector that encodes the second-order interactions between users/items and their attributes in the embedding space.

It is worth pointing out that although pairwise pooling models the correlation between each pair of features, it can be efficiently computed in linear time --- the same time complexity with average/max pooling. To show the linear time complexity of evaluating pairwise pooling, we reformulate Eqn.\eqref{equ:user-bilinear-pooling} as,
\begin{gather}
\Mat{p}_{u}=\frac{1}{2}\left[(\Mat{u}+\sum_{t=1}^{V_{u}}\Mat{g}_{t}^{u})\odot(\Mat{u}+\sum_{t=1}^{V_{u}}\Mat{g}_{t}^{u})-\Mat{u}\odot\Mat{u}-\sum_{t=1}^{V_{u}}\Mat{g}_{t}^{u}\odot\Mat{g}_{t}^{u}\right],
\end{gather}
which can be computed in $O(K V_{u})$ time. This is a very appealing property, meaning that the benefit of pairwise pooling in modelling all pair-wise correlations does not involve any additional cost, as compared to the average pooling that does not model any correlation between input features.
%Furthermore, factorization machine (FM)~\cite{DBLP:conf/icdm/Rendle10,DBLP:conf/sigir/RendleGFS11,DBLP:journals/tist/Rendle12} can be interpreted as one specialization of our bilinear pooling.

\textbf{Hidden Layers:}~Above the pairwise pooling is a stack of full connected layers, which enable us to capture the nonlinear and higher-order correlations among users, items, and attributes.
Inspired by the neural network view of matrix factorization (\cf Figure~\ref{fig:mf_limit}), we first merge user representation $\Mat{p}_{u}$ and item representation $\Mat{q}_i$ with an element-wise product, which models the two-way interaction between $u$ and $i$. We then place a multi-layer perceptron (MLP) above the element-wise product. Formally, the hidden layers are defined as:
\begin{gather}\label{equ:fully-connected-layer}
\begin{cases}
\Mat{e}_{1}=\sigma_{1}(\Mat{W}_{1}(\Mat{p}_{u}\odot\Mat{q}_{i})+\Mat{b}_{1})\\
\Mat{e}_{2}=\sigma_{2}(\Mat{W}_{2}\Mat{e}_{1}+\Mat{b}_{2})\\
\cdots\cdots\\
\Mat{e}_{L}=\sigma_{L}(\Mat{W}_{L}\Mat{e}_{L-1}+\Mat{b}_{L})\\
\end{cases},
\end{gather}
where $\Mat{W}_{l}$, $\Mat{b}_{l}$, $\sigma_{l}$, and $\Mat{e}_{l}$ denote the weight matrix, bias vector, activation function, and output vector of the $l$-th hidden layers, respectively. As for the activation function in each hidden layer, we opt for Rectifier (ReLU) unit, which is more biologically plausible and proven to be non-saturated.
%to learn higher-order user-item interactions in a non-linear way.
Regarding the structure of hidden layers, common choices include the tower~\cite{heneural,DBLP:conf/recsys/CovingtonAS16}, constant, and diamond, among others. In this work, we simply set all hidden layers have the same size, leaving the further tuning of the deep structure as the future work.

\textbf{Prediction Layer:}~At last, the output vector of the last hidden layer $\Mat{e}_{L}$ is transformed to the prediction score:
\begin{gather}\label{equ:prediction-layer}
\hat{y}_{ui}=\Trans{\Mat{w}}\Mat{e}_{L},
\end{gather}
where $\Mat{w}$ represents the weight vector of the prediction layer.

Note that we have recently proposed a neural factorization machine (NFM) model~\cite{he2017neural}, which similarly uses a pairwise pooling operation to model the interaction among features. We point out that the main architecture difference is in our separated treatment of the user and item channel, where each channel can essentially be seen as an application of NFM on the user/item ID and attributes.

\subsection{Learning of Social Domain}

With the above neural collaborative ranking solution, we obtain an attribute-aware representation $\Mat{p}_u$ and $\Mat{q}_i$ for each user and item, respectively.
To predict the affinity score of a social user to an item of the information domain, we need to also learn an representation for the social user in the same latent space of the information domain. We achieve this goal by propagating $\Mat{p}_u$ from bridge users to representations for non-bridge users of the social domain. The intuition for such representation propagation is that, if two users are strongly connected (\eg close friends with frequent interactions), it is likely that they have the similar preference on items; as such, they should have similar representations in the latent space. This suits well the paradigm of graph regularization~\cite{DBLP:conf/cikm/HeCKC15,DBLP:journals/tkde/WangFHLW17,DBLP:journals/tkde/WangFHTW16,fuli2017computational} (\aka semi-supervised learning on graph), which has two components:
%By using the representation of bridge users (i.e., partial users of the information domain) to regularize the learning, we

%Through the neural model above, the users and items in the information domain have been projected into the same semantic space with high engagements. With the representations of bridge users, we can guide the embedding learning of the social users in SNSs, which should be guaranteed to be in the same space as the items. Towards this end, we introduce two regularization terms, \emph{fitting regularizer} and \emph{smoothness regularizer} into our social network embedding phase as,

%\begin{gather}
%\Lapl_{u}=\mu\theta(\Set{U})+\theta(\Set{U}_{2}),
%\end{gather}
%where $\mu$ is a positive parameter to capture the tradeoff between the fitting regularizer $\theta(\Set{U})$ and the normalized graph Laplacain $\theta(\Set{U}_{2})$.

\textbf{Smoothness:}~The smoothness constraint implies the structural consistency --- the nearby vertices of a graph should not vary much in their representations. Enforcing smoothness constraint in our context of social domain learning will propagate a user's representation to her neighbors, such that when a steady state reaches, all vertices should have been placed in the same latent space. The objective function for smoothness constraint is defined as:
\begin{gather}\label{equ:smoothness}
\theta(\Set{U}_{2})=\frac{1}{2}\sum_{u',u''\in\Set{U}_{2}}s_{u'u''}\norm{\frac{\Mat{p}_{u'}}{\sqrt{d_{u'}}}-\frac{\Mat{p}_{u''}}{\sqrt{d_{u''}}}}^{2},
\end{gather}
where $s_{u'u''}$ denotes the strength of social connection between $u'$ and $u''$, and $d_{u'}$ (or $d_{u''}$) denotes the outdegree of $u'$ (or $u''$) for normalization purpose. It is worth noting that the use of normalization is the key difference with the social regularization used by \cite{DBLP:conf/wsdm/MaZLLK11,zhao2016user}, which does not apply any normalization on the smoothness constraint. As pointed out by He \etal~\cite{DBLP:conf/cikm/HeCKC15}, the use of normalization helps to suppress the impact of popular vertices, which can lead to more effective propagation. We empirically verify this point in Section 4.3.

%Inspired by previous studies~\cite{DBLP:conf/kdd/Bressan0PRT16,DBLP:series/synthesis/2016NieSC}, we assume that, the social users hold a better-than-average chance in adopting items rated by their friends. In the light of this, we argue the users in the same social circle tend to have the similar embedding, which hold the similar correlations with the items. We devise the social regularizer on user-user social connections as follow,

%Moreover, we can rewrite the Eqn.\eqref{equ:smoothness} in the form of graph Laplacian as,
%\begin{gather}
%\theta(\Set{U}_{2})=\frac{1}{2}\Trace{\Trans{\Mat{P}}(\Mat{I}-\Mat{D}^{-\frac{1}{2}}\Mat{S}\Mat{D}^{-\frac{1}{2}})\Mat{P}},
%\end{gather}
%where $\Mat{P}=[\Mat{p}_{1},\cdots,\Mat{p}_{M_{2}}]\in\Space{R}^{K\times M_{2}}$ represents the representation matrix of social users; $\Mat{S}\in\Space{R}^{M_{2}\times M_{2}}$ denotes the affinity matrix, with its $(u,u')$-th entry as $s_{uu'}$; $\Mat{D}\in\Space{R}^{M_{2}\times M_{2}}$ denotes the diagonal degree matrix with its $(u,u)$-th entry as $d_{u}$; $\Mat{I}$ is an identity matrix. Whereinto, this regularization is known as \emph{smoothness regularizer} or \emph{normalized Laplacian}, which not only preserves the first-order proximity (i.e., the pairwise similarity between any two users) in the social network via the squared loss~\cite{DBLP:conf/cikm/HeCKC15}, but also prevent the embedding from being dominated by popular social neighbors via the normalization terms~\cite{DBLP:conf/www/BalujaSSJYKRA08}.

\textbf{Fitting:}
~The fitting constraint implies the latent space consistency across two domains --- the bridge users' representations should be invariant and act as the anchors across domains. Towards this end, we encourage the two representations of the same bridge users to be close to each other. The objective function for fitting constraint is defined as,
\begin{gather}\label{equ:fitting}
\theta(\Set{U})=\frac{1}{2}\sum_{u'\in\Set{U}}\norm{\Mat{p}_{u'}-\Mat{p}_{u'}^{(0)}}^{2},
\end{gather}
where for each bridge user $u'$, $\Mat{p}_{u'}$ (or $\Mat{p}_{u'}^{(0)}$) is her representation of the SNS (or information domain). As such, the fitting constraint essentially acts as the bridges connecting the two latent spaces.

Lastly, we combine the smoothness constraint with the fitting constraint and obtain the objective function of the social domain learning as,
\begin{gather}
\Lapl_{S}=\theta(\Set{U}_{2})+\mu\theta(\Set{U}),
\end{gather}
where $\mu$ is a positive parameter to control the tradeoff between two constraints.

\subsubsection{\textbf{Prediction for Social Users}}
With the representations of social users and items (\ie $\Mat{p}_{u'}$ and $\Mat{q}_{i}$) at hand, we can feed them into the fully connected layers as Eqn.\eqref{equ:fully-connected-layer} shows and utilize the prediction layer as Eqn.\eqref{equ:prediction-layer} displays. At last, we can obtain the predicted preference $\widehat{y}_{u'i}$, as follows,
\begin{gather}
\begin{cases}
\Mat{e}_{1}=\sigma_{1}(\Mat{W}_{1}(\Mat{p}_{u'}\odot\Mat{q}_{i})+\Mat{b}_{1})\\
\cdots\cdots\\
\Mat{e}_{L}=\sigma_{L}(\Mat{W}_{L}\Mat{e}_{L-1}+\Mat{b}_{L})\\
\widehat{y}_{u'i}=\Trans{\Mat{w}}\Mat{e}_{L}
\end{cases}.
\end{gather}

\subsection{Training}
We adopt the alternative optimization strategy on Eqn.\eqref{equ:framework} since it can emphasize exclusive characteristics within individual domains. In the information domain, we employ stochastic gradient descent SGD) to train the attribute-aware NSCR in the mini-batch mode and update the corresponding model parameters. In particular, we first sample a batch of observed user-item interactions $(u, i)$ and adopt negative sampling~\cite{heneural} to randomly select an unobserved item $j$ for each $(u, i)$. We then generate a triplet $(u,i,j)$. Following that, we take a gradient step to optimize the loss function $\Lapl_{I}$ in Eqn.\eqref{equ:information-obj}. As such, we obtain the enhanced representations of users. In the SNS, we feed the enhanced representations of bridge users into our graph Laplacian to update all representations of social users. Towards this end , we can simplify the derivative of $\Lapl_{S}$ regarding user representation $\Mat{P}$ and then obtain the close-form solution as,
\begin{gather}\label{equ:social-training}
\Mat{P}=\frac{\mu}{1+\mu}\left(\Mat{I}-\frac{1}{1+\mu}\Mat{D}^{-\frac{1}{2}}\Mat{S}\Mat{D}^{-\frac{1}{2}}\right)^{-1}\Mat{P}^{(0)},
\end{gather}
where $\Mat{P}^{(0)}$ is the embedding of social users, which includes the updated representations of bridge users from NSCR part; $\Mat{S}$ and $\Mat{D}$ are the similarity matrix and diagonal degree matrix of social users, respectively, whereinto $S_{u'u''}=s_{u'u''}$ and $D_{u'u'}=d_{u'}$. Thereafter, we view the newly updated representations of bridge users as the next initialization for the bridge users in NSCR. We repeat the above procedures to approximate the model parameter set $\Theta$. As for the regularization term in Eqn.\eqref{equ:framework}, we omit it since we utilize \emph{dropout} technique in neural network modeling to avoid overfitting.

%As we jointly learn the neural CR and social network embedding, we adopt the alternating strategy of the optimization. In particular, we employ stochastic gradient descent (SGD)~\cite{bottou2010large} to train the neural collaborative ranking in the mini-batch mode and update the corresponding model parameters. We first sample a batch of observed user-item interactions $(u, i)$ and adopt negative sampling~\cite{DBLP:conf/nips/MikolovSCCD13} to randomly select an unobserved item $j$ for each $(u, i)$. We then generate a triplet $(u,i,j)$. Following that, we take a gradient step to optimize the loss function $\Lapl_{s}$ of our neural CR model. Thereafter, to reconstruct the social network, we can simply take the derivative of $\Lapl_{u}$ regarding user representation $\Mat{P}$ as,
%\begin{gather}\label{equ:social-training}
%\frac{\partial \Lapl_{u}}{\partial \Mat{P}}=\frac{\mu}{1+\mu}\left(\Mat{I}-\frac{1}{1+\mu}\Mat{D}^{-\frac{1}{2}}\Mat{S}\Mat{D}^{-\frac{1}{2}}\right)^{-1}\Mat{P}^{(0)},
%\end{gather}
%where $\Mat{P}^{0}$ is the embedding of social users, which includes the updated representations of bridge users from neural CR model. We use the learned embedding $\Mat{P}$ as initialization for joint training. We repeat the above procedures to approximate the model parameter set $\Theta$. As for the regularization term in Eqn.\eqref{equ:framework}, we omit it in this work, since we utilize \emph{dropout} technique in neural network modeling to avoid overfitting.

\textbf{Dropout:}~Dropout is an effective solution to prevent deep neural networks from overfitting. The idea is to randomly drop part of neurons during training. As such, only part of the model parameters, which contribute to the final ranking, will be updated. In our neural CR model, we propose to adopt dropout on the pairwise pooling layer. In particular, we randomly drop $\rho$ of $\Mat{p}_{u}$ and $\Mat{q}_{i}$, whereinto $\rho$ is the dropout ratio. Analogous to the pooling layer, we also conduct dropout on each hidden layer.

\section{Experiments}
\label{sec:experiments}
To comprehensively evaluate our proposed method, we conducted experiments to answer the following research questions:
\begin{itemize}[leftmargin=*]
	\item\textbf{RQ1:}~Can our NSCR approach outperform the state-of-the-art recommendation methods for the new cross-domain social recommendation task?
	\item\textbf{RQ2:}~How do different hyper-parameter settings (\eg the dropout ratio and tradeoff parameters) affect NSCR?
	\item\textbf{RQ3:}~Are deeper hidden layers helpful for learning from user-item interaction data and improving the performance of NSCR?
	%? Does cross-domain social recommendation benefit from the neural network structure?
\end{itemize}
%In the followings, we first describe the data construction process and experimental settings. We  then answer the above three research questions one by one.

\subsection{Data Description}

	To the best of our knowledge, there is no available public benchmark dataset that
	fits the task of cross-domain social recommendation. As such, we constructed the datasets by ourselves. We treated Trip.com as the information domain, Facebook and Twitter as the social domains. In Trip.com, we initially compiled $6,532$ active users, who had at least $5$ ratings over $2,952$ items (\eg \emph{gardens by the bay} in Singapore and \emph{eiffel tower} in Pairs). We transformed their $93,998$ ratings into binary implicit feedback as ground truth, indicating whether the user has rated the item.
	Moreover, we collected $19$ general categories regarding the travel mode (\eg \emph{adventure travel}, \emph{business travel}, and \emph{nightlife}) and used them as the attributes of users and items.
	%Analogous to user-item interactions, the binary belonging is denoted as $0$ or $1$ indicating whether the user/item has the attribute.
	Subsequently, we parsed the users' profiles to identify their aligned accounts in Facebook and Twitter, inspired by the methods in~\cite{DBLP:series/synthesis/2016NieSC,DBLP:conf/sigir/SongNZAC15}.
	We obtained $858$ and $502$ bridge users for Facebook and Twitter, respectively. Thereafter, we crawled the public friends or followers of each bridge user to reconstruct the social networks, resulting in $177,042$ Facebook users and $106,049$ Twitter users.
	However, the original social data are highly sparse, where most non-bridge users have only one friend, making it ineffective to propagate users' preferences.
	%It is hence difficult to effectively propagate the social influence and evaluate the performance of collaborative ranking.
	To ensure the quality of the social data,
	we performed a modest filtering on the data, retraining users with at least two friends.
	%we developed a filter principle to clean the data and retained only users with at least two social connections (friends).
	This results in a subset of the social data that contains $7,233$ Twitter users with $42,494$ social connections and $8,196$ Facebook users with $49,156$ social connections. The statistics of the datasets are summarized in Table~\ref{tab:data-statistics}.
	
	\subsection{Experimental Settings}
	\textbf{Evaluation Protocols:}~Given a social user, each method generates an item ranking list for the user.
	To assess the ranking list, we adopted two popular IR metrics, $AUC$ and $recall$, to measure the quality of preference ranking and top-$N$ recommendation.
	\begin{itemize}[leftmargin=*]
		%XN： revise to be more professional.
		\item\textbf{AUC:}~Area under the curve (AUC)~\cite{DBLP:conf/uai/RendleFGS09,DBLP:conf/www/HuCXCGZ13} measures the probability that a recommender system ranks a positive user-item interaction higher than negative ones:
		\begin{gather}
		AUC = \frac{\sum_{i\in\Set{I}_{u}^{+}}\sum_{j\in\Set{I}_{u}^{-}}\delta(\widehat{y}_{uij}>0)}{|\Set{I}_{u}^{+}||\Set{I}_{u}^{-}|},
		\end{gather}
		where $\Set{I}_{u}^{+}=\{i|y_{ui}=1\}$ and $\Set{I}_{u}^{-}=\{j|y_{uj}=0\}$ denote the sets of relevant (observed) item $i$ and irrelevant (unobserved) item $j$ for user $u$, respectively; and $\delta$ is the count function returning $1$ if $\widehat{y}_{uij}>0$ and $0$ otherwise. Below we report the averaged AUC for all testing users.
		\item\textbf{R@$\Mat{K}$:}~Recall@$K$ considers the relevant items within the top $K$ positions of the ranking list. A higher recall with lower $K$ indicates a better recommender system, which can be defined as,
		\begin{gather}
		R@K = \frac{|\Set{I}_{u}^{+}\cap\Set{R}_{u}|}{|\Set{I}_{u}^{+}|},
		\end{gather}
		where $\Set{R}_{u}$ denotes the set of the top-$K$ ranked items for the given user $u$. Analogous to AUC, we report the average $R@5$ for all testing users.
	\end{itemize}
	
	\noindent By learning representations for social users and information-domain items together, our NSCR is capable of recommending items for both bridge and non-bridge users.
	However, due to the limitation of our static datasets, it is difficult for us to evaluate the recommendation quality for non-bridge users, since they have no interaction on the information-domain items.
	As such, we rely on the bridge users for evaluating the performance.
	Following the common practice in evaluating a recommender algorithm~\cite{heneural,DBLP:conf/uai/RendleFGS09},
	we holdout the latest $20\%$ interactions of a bridge user as the test set.
	To tune hyper-parameters, we further randomly holdout $20\%$ interactions from a bridge user's training data as the validation set. We feed the remaining bridge users, all the non-bridge users in SNSs, and the remaining user-item interactions in the information domains into our framework for training.
	
	\begin{table}[t]
		\centering
		\caption{Statistics of the complied datasets. The social user set includes the bridge users.}
		\vspace{-5pt}
		\label{tab:data-statistics}
		\resizebox{0.46\textwidth}{!}{
			\begin{tabular}{|l|c|c|c|}
				\hline
				\textbf{Information Domain} & \textbf{User\#}  & \textbf{Item\#}  & \textbf{Interaction\#}         \\ \hline
				Trip.com       & $6,532$                & $2,952$                & $93,998$                       \\ \hline\hline
				\textbf{SNSs}   & \textbf{Bridge User\#} & \textbf{Social User\#} & \textbf{Social Connection\#}   \\ \hline
				Twitter        & $502$                  & $7,233$                & $42,494$                       \\ %\hline
				Facebook       & $858$                & $8,196$                & $49,156$                       \\ \hline
		\end{tabular}}
		\vspace{-1em}
	\end{table}

	\textbf{Baselines:}~To justify the effectiveness of our proposal, we study the performance of the following methods:
	\begin{itemize}[leftmargin=*]
		\item\textbf{ItemPop:}~This method ranks items base on their popularity, as judged by the number of interactions. It is a non-personalized method that benchmarks the performance of a personalized system~\cite{DBLP:conf/uai/RendleFGS09}.
		\item\textbf{MF:}~This is the standard matrix factorization model that leverages only user--item interactions of the information domain for recommendation~(\cf Eqn.\eqref{equ:mf}).
		\item\textbf{SFM:}~Factorization machine~\cite{DBLP:conf/icdm/Rendle10} is a generic factorization model that is designed for recommendation with side information.
		We construct the input feature vector by using one-hot encoding on the ID and attributes of users and items.
		To adjust FM for modelling social relations, we further plug a (bridge) user's friends into the input feature vector, dubbed this enhanced model as Social-aware FM (SFM).
		%To explore the efficacy of attributes, we also evaluate a variant SFM-a that removes attribute information from the input feature vector.
		\item\textbf{SR:}~This~\cite{DBLP:conf/wsdm/MaZLLK11} is a state-of-the-art factorization method for social recommendation.
		It leverages social relations to regularize the latent vectors of friends to be similar.
		To incorporate attributes into their method, we adjust the similarity of two users based on their attribute sets, which leads to better performance.
		%\item\textbf{SFM/SFM-a:}~FM is a powerful factorization model with feature engineering. We modified the basic model~\cite{DBLP:conf/uai/RendleFGS09} by incorporating the attribute variables, and then applied hidden layers as the same as NSCR, named SFM. For each user-item interaction, we concatenated all variables related to the user as the input, including her friends, user-specific, and item-specific attributes. To verify the efficacy of attributes, we removed attribute information and degraded SFM to SFM-a.
		%\item\textbf{SR/SR-a:}~SR~\cite{DBLP:conf/wsdm/MaZLLK11} is a state-of-the-art factorization method with social regularization for item recommendation. It optimizes the squared loss of observed user-item interaction and the social regularization based on similarity of friends. To make it suitable for our task, we modified its squared loss function with the pairwise learning loss as Eqn.\eqref{equ:ranking-loss} and added a stack of hidden layers similar to NSCR. Furthermore, we employed the Jaccard coefficient on the pairs of a user' attribute set as the additional user similarity (connections). Analogous to SFM-a, we removed attribute information from SR and denoted it as SR-a.
	\end{itemize}
	
	\noindent Note that for all model-based methods, we optimize them with the same pair-wise ranking function of Eqn.\eqref{equ:information-obj} for a fair comparison on the model's expressiveness.
	To explore the efficacy of attributes, we further explore variants that remove attribute modelling from SFM, SR, and NSCR, named as SFM-a, SR-a, and NSCR-a, respectively.
	%For a fair comparison, we added a stack of hidden layers on the embedding of the user-item interactions of FM/FM-a and SR/SR-a.
	
	\textbf{Parameter Settings:}~We implemented our proposed framework on the basis of Tensorflow\footnote{\url{https://www.tensorflow.org}.}, which will be made publicly available, as well as our datasets. For all the neural methods, we randomly initialized model parameters with a Gaussian distribution, whereinto the mean and standard deviation is $0$ and $0.1$, respectively. The mini-batch size and learning rate for all methods was searched in $[128,256,512,1024]$ and $[0.0001,0.0005,0.001,0.05,0.1]$, respectively. We selected Adagrad as the optimizer. Moreover, we empirically set the size of hidden layer same as the embedding size (the dimension of the latent factor) and the activation function as ReLU. Without special mention, we employed two hidden layers for all the neural methods, including SFM, SR, and NSCR. We randomly generated ten different initializations and feed them into our NSCR. For other competitors, the initialization procedure is analogous to ensure the fair comparison. Thereafter, we performed paired t-test between our model and each of baselines over $10$-round results.

\subsection{Performance Comparison (RQ1)}
We first compare the recommendation performance of all the methods. We then purpose to justify how the social modelling and the attribute modelling affect the recommendation performance.

\begin{table}[t]
	\centering
	\caption{Performance comparison between all the methods, when the embedding size$=64$ and signiﬁcance test is based on AUC.}
	\vspace{-5pt}
	\label{tab:p-value}
	\resizebox{0.46\textwidth}{!}{\begin{tabular}{|c|c|c|c|c|c|c|}
			\hline
			\textbf{Datasets} & \multicolumn{3}{c|}{\textbf{Twitter-Trip}}                       & \multicolumn{3}{c|}{\textbf{Facebook-Trip}}                      \\ \hline
			\textbf{Methods}  & \textbf{AUC}   & \textbf{R@$\Mat{5}$} & \textbf{$\Mat{p}$-value} & \textbf{AUC}   & \textbf{R@$\Mat{5}$} & \textbf{$\Mat{p}$-value} \\ \hline\hline
			\textbf{ItemPop}  & $0.7193$       & $0.0164$             & $3e$-$5$                         & $0.7439$       & $0.0249$             & $8e$-$6$                         \\ \hline
			\textbf{MF}       & $0.8285$       & $0.0375$             & $3e$-$4$                         & $0.8596$       & $0.0821$             & $1e$-$4$                         \\ \hline
			\textbf{SFM}      & $0.8832$       & $0.0492$             & $2e$-$3$                         & $0.8908$       & $0.0856$             & $1e$-$3$                         \\ \hline
			\textbf{SR}       & $0.9013$       & $0.0747$             & $9e$-$3$                         & $0.9267$       & $0.1433$             & $4e$-$2$                         \\ \hline
			\textbf{NSCR}     & $\Mat{0.9222}$ & $\Mat{0.0807}$       & -                        & $\Mat{0.9390}$ & $\Mat{0.1466}$       & -                        \\ \hline
		\end{tabular}}
		\vspace{-1em}
	\end{table}
	
	\begin{figure*}[t]
		\centering
		\subfigure[AUC on Twitter-Trip]{
			\includegraphics[width=0.235\textwidth]{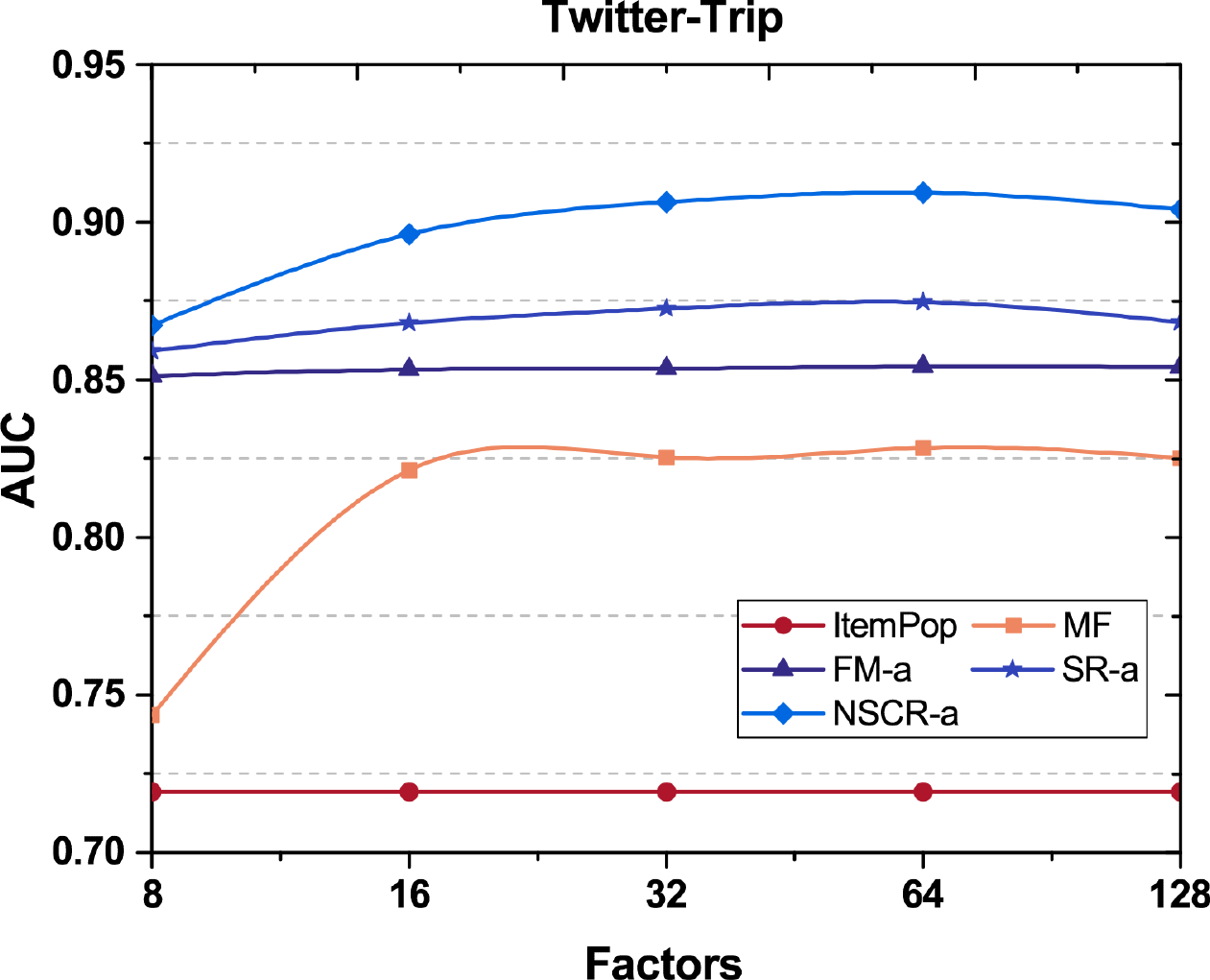}
			\label{fig:without-attr-tw-auc}}
		\subfigure[R@5 on Twitter-Trip]{
			\includegraphics[width=0.235\textwidth]{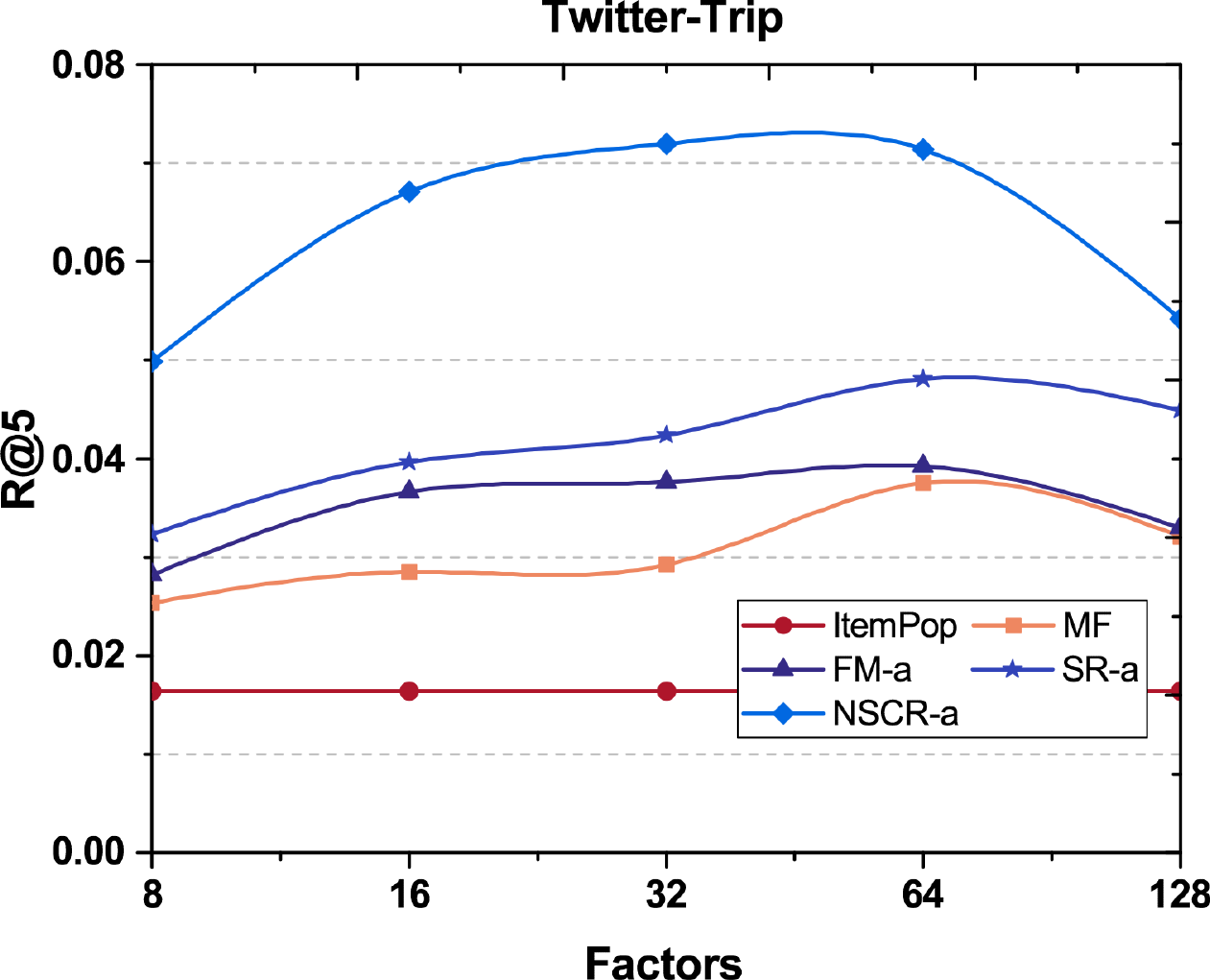}
			\label{fig:without-attr-tw-auc}}
		\subfigure[AUC on Facebook-Trip]{
			\includegraphics[width=0.235\textwidth]{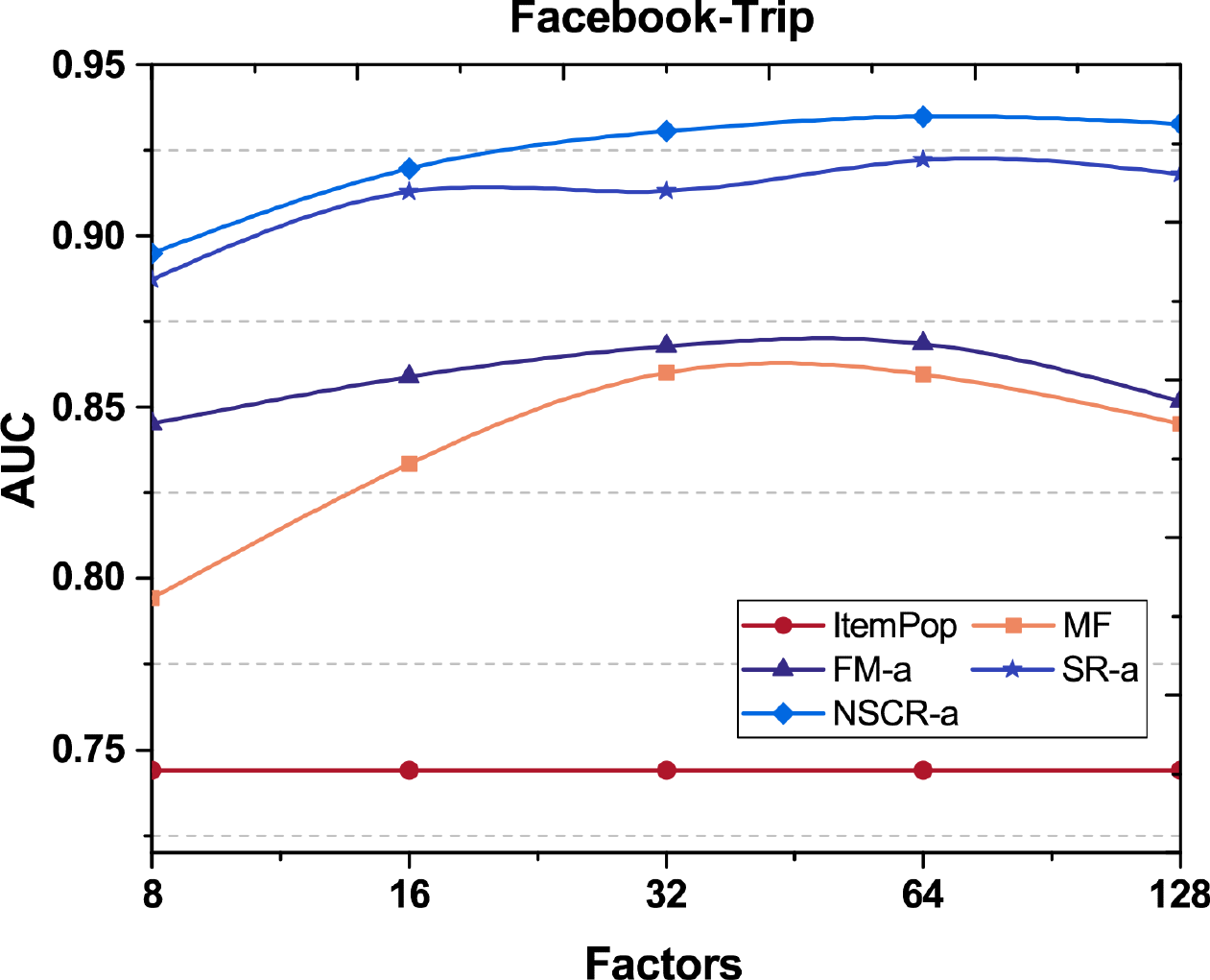}
			\label{fig:without-attr-fb-auc}}
		\subfigure[R@5 on Facebook-Trip]{
			\includegraphics[width=0.235\textwidth]{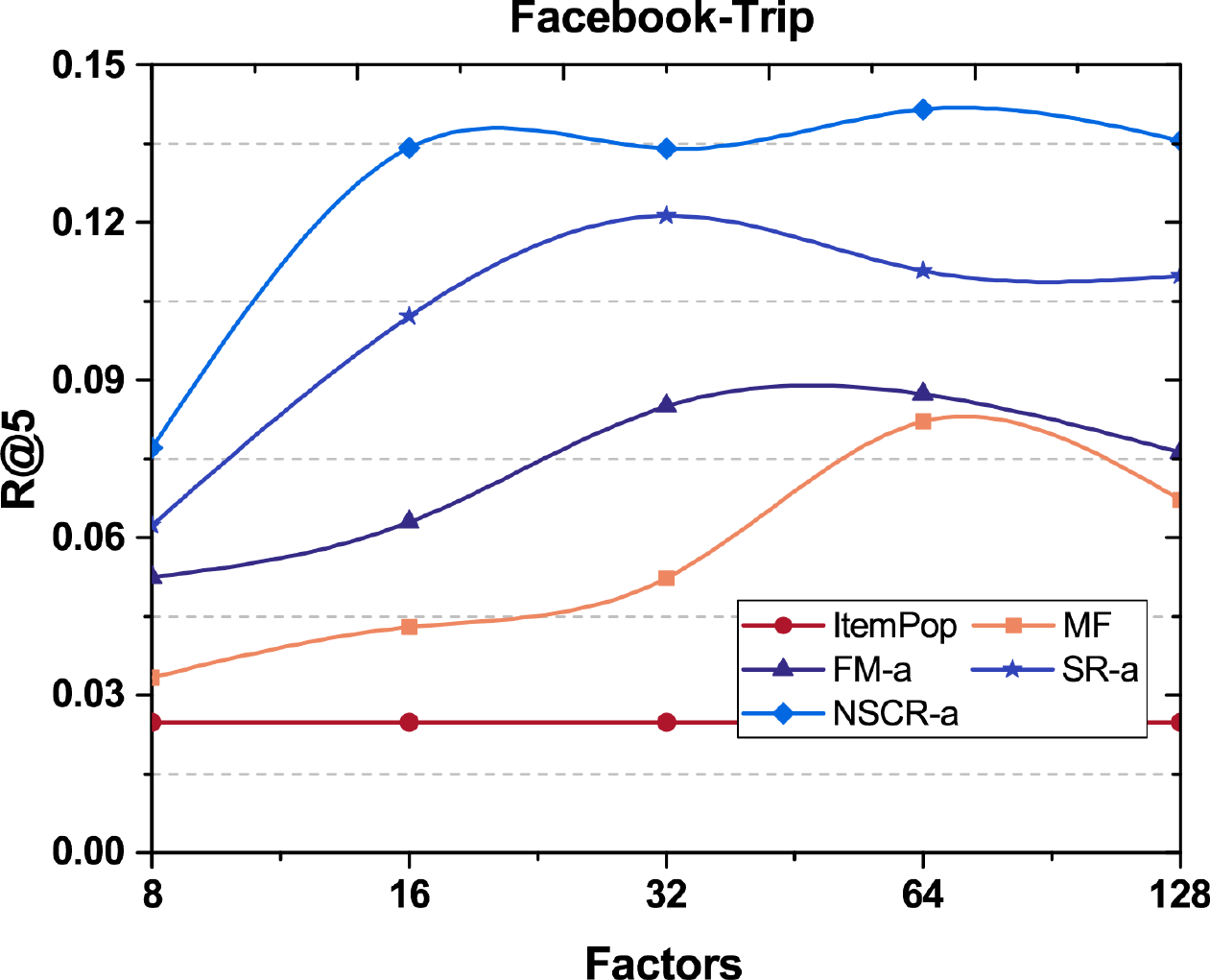}
			\label{fig:without-attr-fb-auc}}
		\vspace{-10pt}
		\caption{Performance comparison of AUC and R@$5$ \wrt the embedding size on Twitter-Trip and Facebook-Trip datasets.}
		\vspace{-1em}
		\label{fig:performance-social}
	\end{figure*}
	
	\begin{figure*}[h]
		\centering
		\subfigure[AUC on Twitter-Trip]{
			\includegraphics[width=0.235\textwidth]{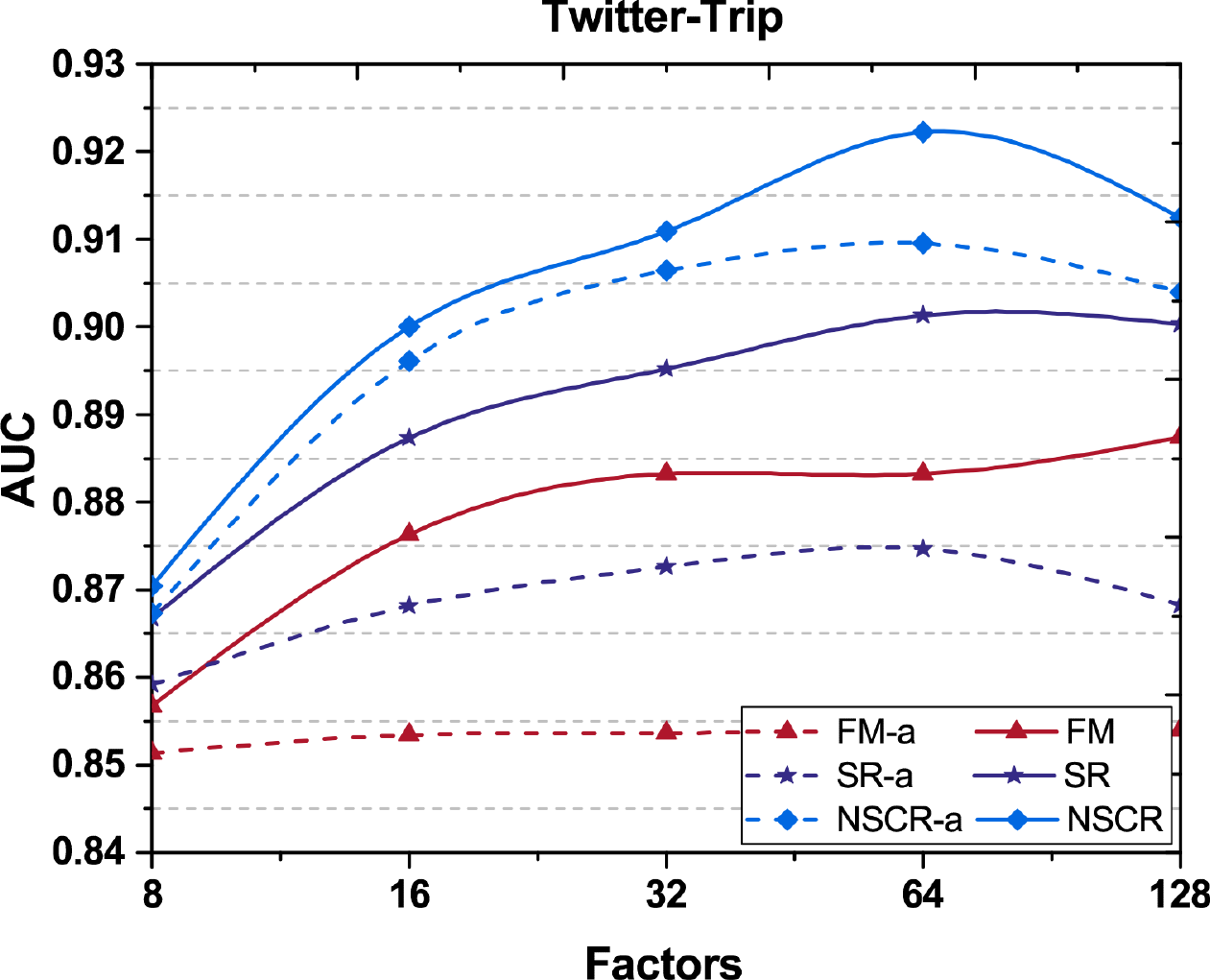}
			\label{fig:with-attr-tw-auc}}
		\subfigure[R@5 on Twitter-Trip]{
			\includegraphics[width=0.235\textwidth]{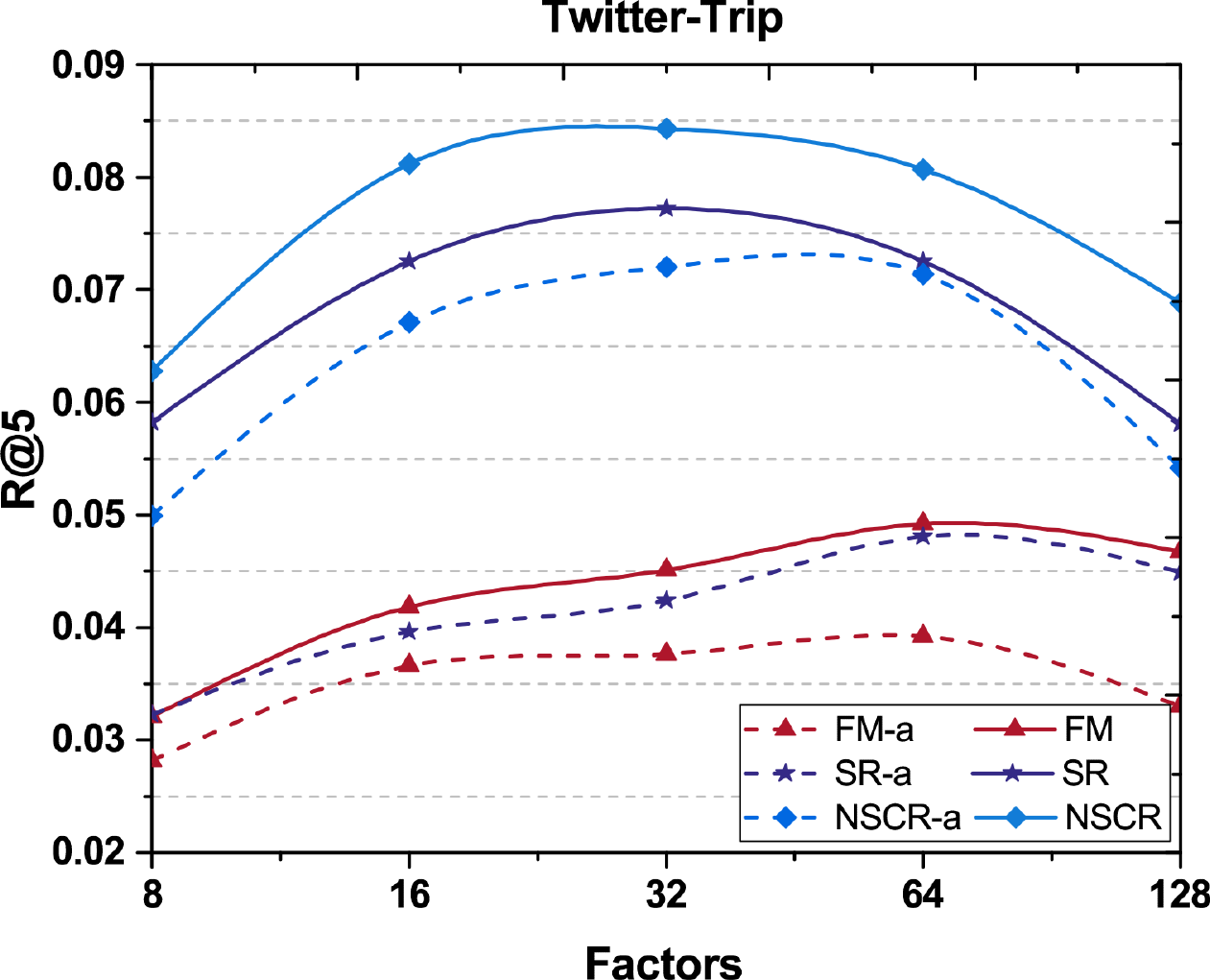}
			\label{fig:with-attr-tw-auc}}
		\subfigure[AUC on Facebook-Trip]{
			\includegraphics[width=0.235\textwidth]{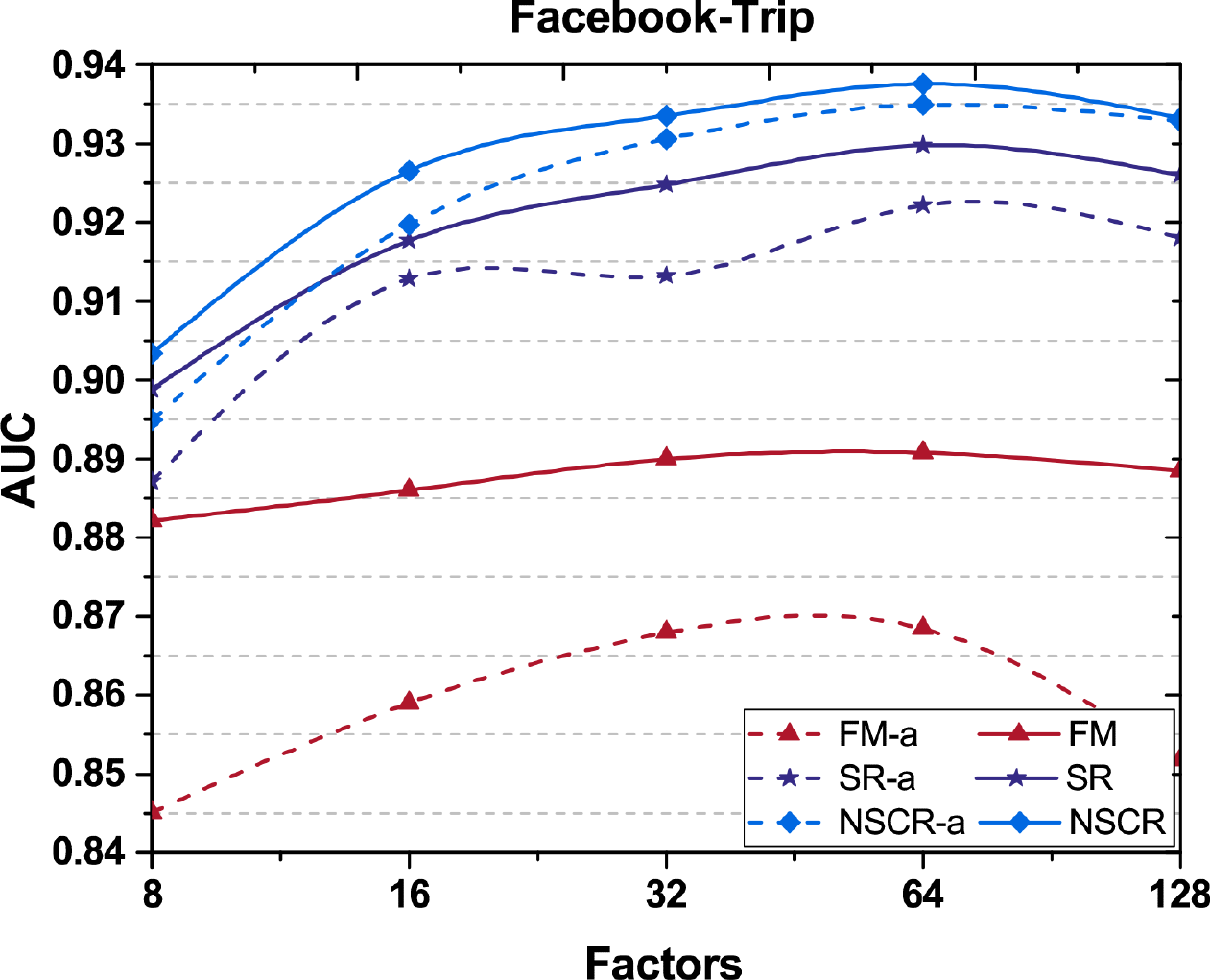}
			\label{fig:with-attr-fb-auc}}
		\subfigure[R@5 on Facebook-Trip]{
			\includegraphics[width=0.235\textwidth]{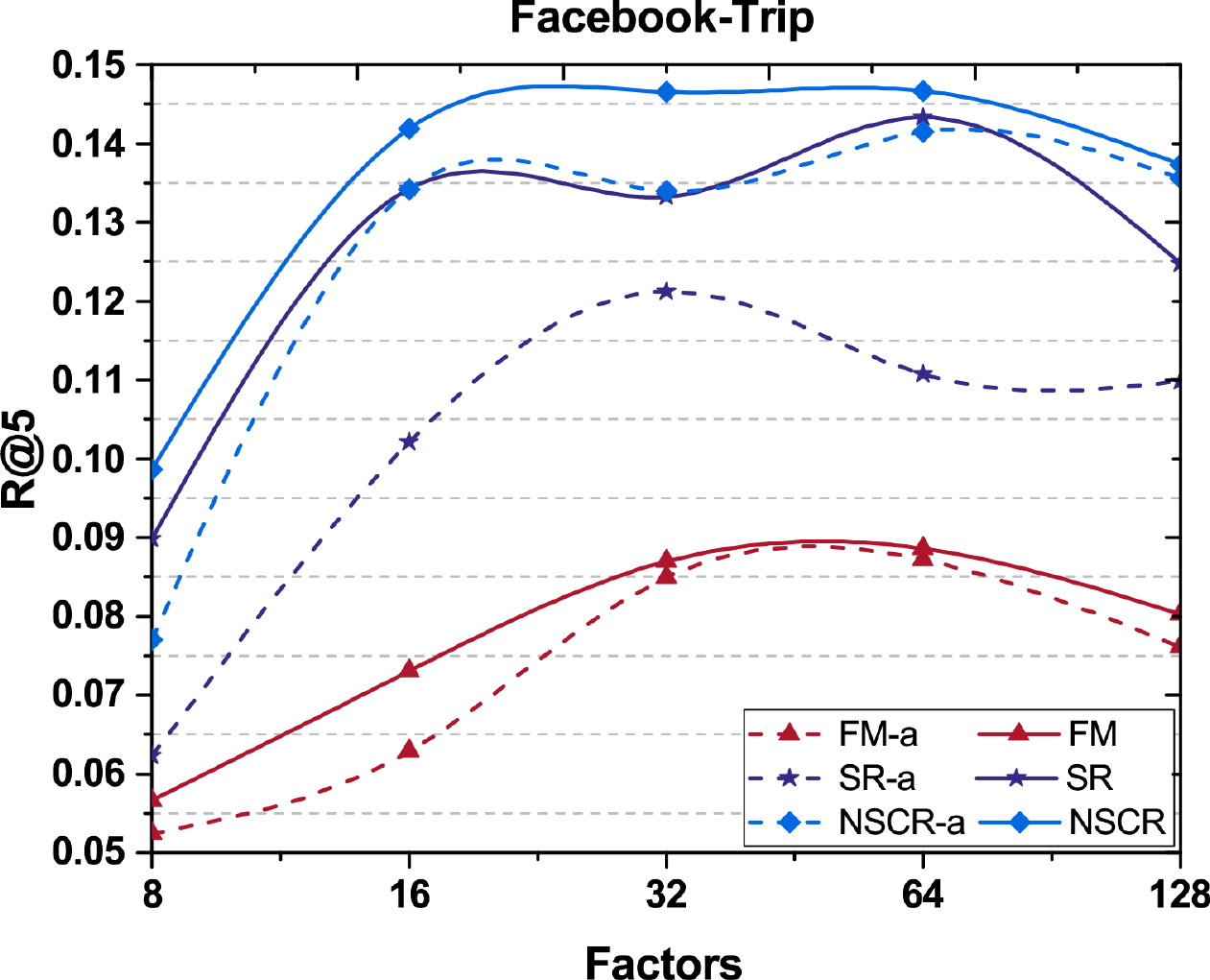}
			\label{fig:with-attr-fb-auc}}
		\vspace{-10pt}
		\caption{Performance comparison of AUC and R@$5$ \wrt the embedding size on Twitter-Trip and Facebook-Trip datasets.}
		\vspace{-1em}
		\label{fig:performance-attribute}
	\end{figure*}
	
	\textbf{Overall Comparison:}~Table~\ref{tab:p-value} displays the performance comparison \wrt AUC and R@$5$ among the recommendation methods on Twitter-Trip and Facebook-Trip datasets, where the embedding size is $64$ for all the methods. We have the following findings:
	\begin{itemize}[leftmargin=*]
		\item ItemPop achieves the worst performance, indicating the necessity of modelling users' personalized preferences, rather than just recommending popular items to users. As for MF, its unsatisfied performance reflects that the independence assumption is insufficient to capture the complex and non-linear structure of user-item interactions.
		\item NSCR substantially outperforms the state-of-the-art methods, SFM and SR. We further conduct one-sample t-tests, verifying that all improvements are statistically significant with $p$-value $<$ $0.05$. It justifies the effectiveness of our proposed framework.
		\item The performance on Twitter-Trip clearly underperforms that of Facebook-Trip. It is reasonable since more bridge users are available in Facebook, which can lead to better embedding learning in SNSs. It again verifies the significance of the bridge users.
	\end{itemize}
	
	\textbf{Effect of Social Modelling:}~To analyze the effect of social modelling, we only consider the variants, SFM-a, SR-a, and NSCR-a. Figure~\ref{fig:performance-social} presents the performance comparison \wrt the number of latent factors on two datasets. We have the following observations.
	\begin{itemize}[leftmargin=*]
		\item ItemPop and MF perform worst since neither of them considers the social connections from SNSs. It highlights the necessity of social modelling in cross-domain social recommendation.
		\item Clearly, NSCR-a significantly outperforms SFM-a and SR-a by a large margin. Formally, in terms of AUC, the relative improvement over SFM-a and SR-a, on average, is $3.19\%$ and $1.01\%$ respectively. While SFM-a considers modelling the social connections, it treats these connections as ordinary features, overlooking the exclusive characteristics of social networks. This leads to the poor expressiveness of the social users' embedding. On the contrary, SR-a and NSCR-a emphasizes the social modelling via the effective social regularization.
		\item Lastly, NSCR-a shows consistent improvements over SR-a, admitting the importance of the normalized graph Laplacian. It again verifies that the normalized graph Laplacian can suppress the popularity of friends and further prevent the social modelling from being dominated by popular social users.
	\end{itemize}
	
	\textbf{Effect of Attribute Modelling:}~As Figure~\ref{fig:performance-attribute} demonstrates, we verify the substantial influence of attribute modelling and the effectiveness of our pairwise pooling operation. Due to the poor performance of ItemPop and MF, they are omitted. Jointly analyzing the performance of all the methods and their variants, we find that,
	\begin{itemize}[leftmargin=*]
		\item For all methods, modelling user/item attributes can achieve significant improvements. By leveraging the similarity of users' attributes, SR enriches the pairwise similarity of any two users and strengthens their connections; meanwhile, SFM can model the correlations of user-attribute, item-attribute, and attribute-attribute, and accordingly enhances the user-item interactions. Benefiting from the pairwise pooling operation, NSCR can encode the second-order interactions between user/item and attributes and boost the representation learning. The significance of attribute is consistent with~\cite{DBLP:conf/mm/ZhangZYYGC13}.
		\item Varying the embedding size, we can see that large embedding may cause overfitting and degrade the performance. In particular, the optimal embedding size is $64$ and $32$ for AUC and R@$5$, respectively. It indicates that the setting of embedding size can effect the expressiveness of our model.
	\end{itemize}
	
	\begin{figure*}[t]
		\centering
		\subfigure[Training Loss]{
			\includegraphics[width=0.235\textwidth]{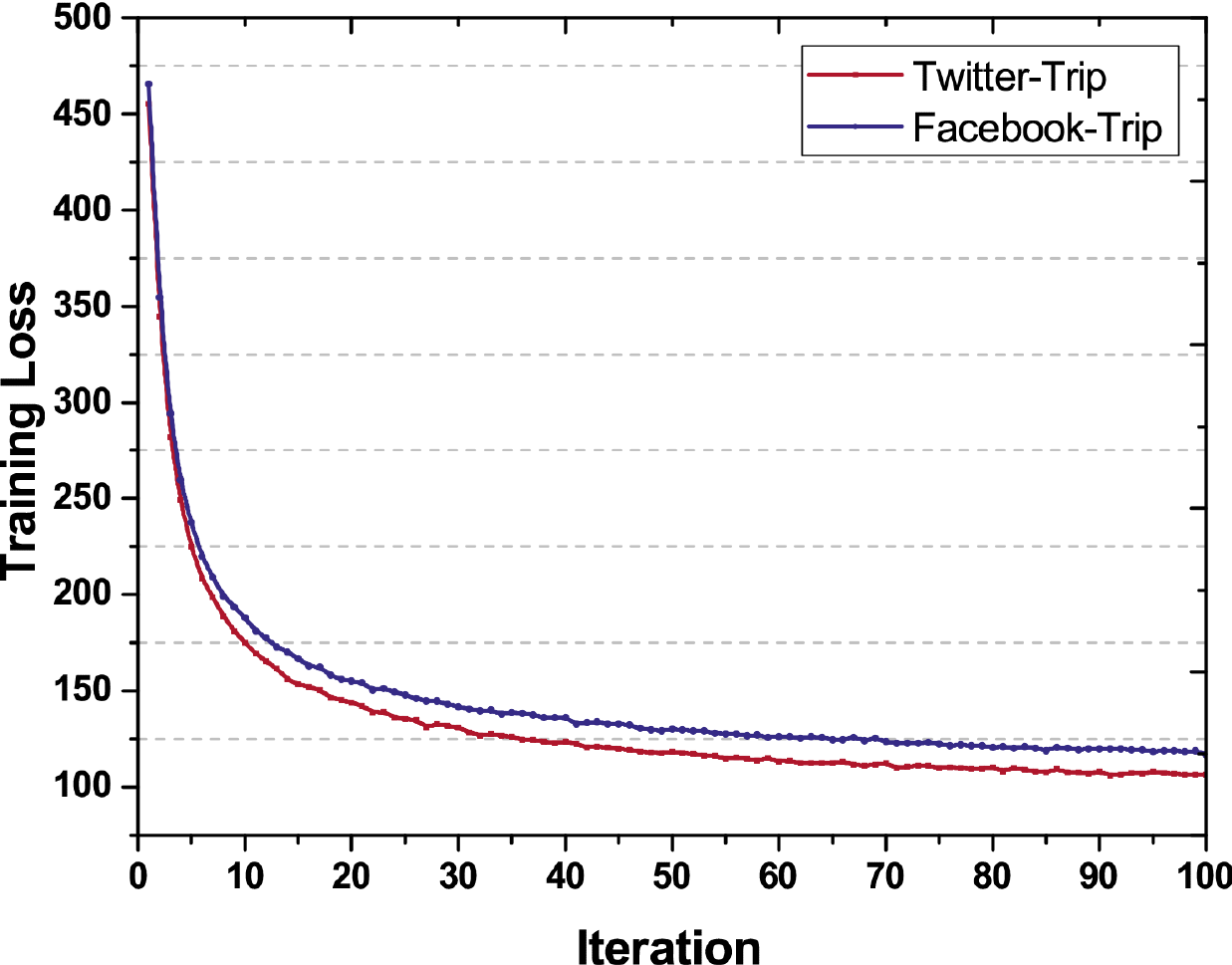}
			\label{fig:training-loss}}
		\subfigure[AUC]{
			\includegraphics[width=0.235\textwidth]{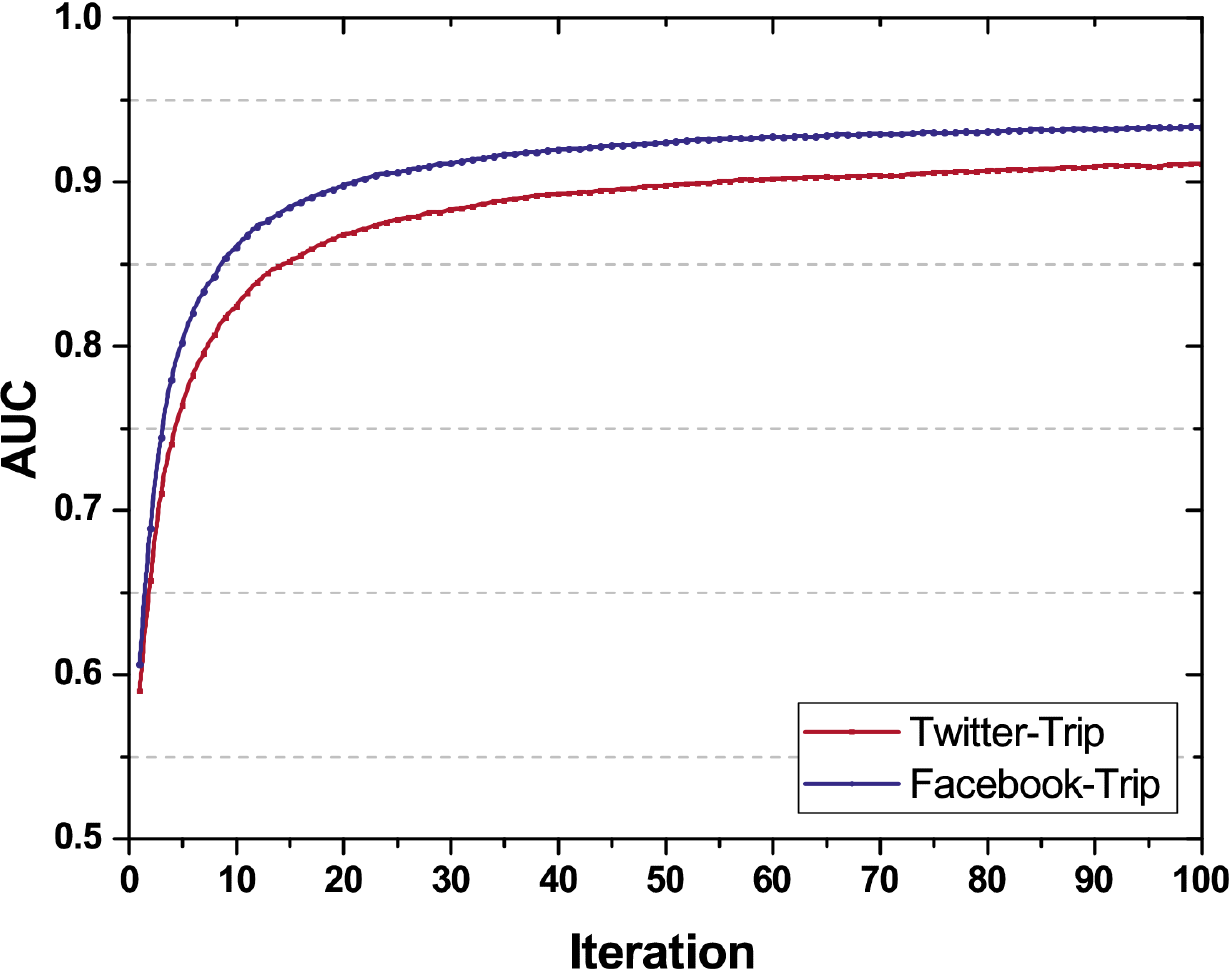}
			\label{fig:auc-sensitivity}}
		\subfigure[R@$5$]{
			\includegraphics[width=0.235\textwidth]{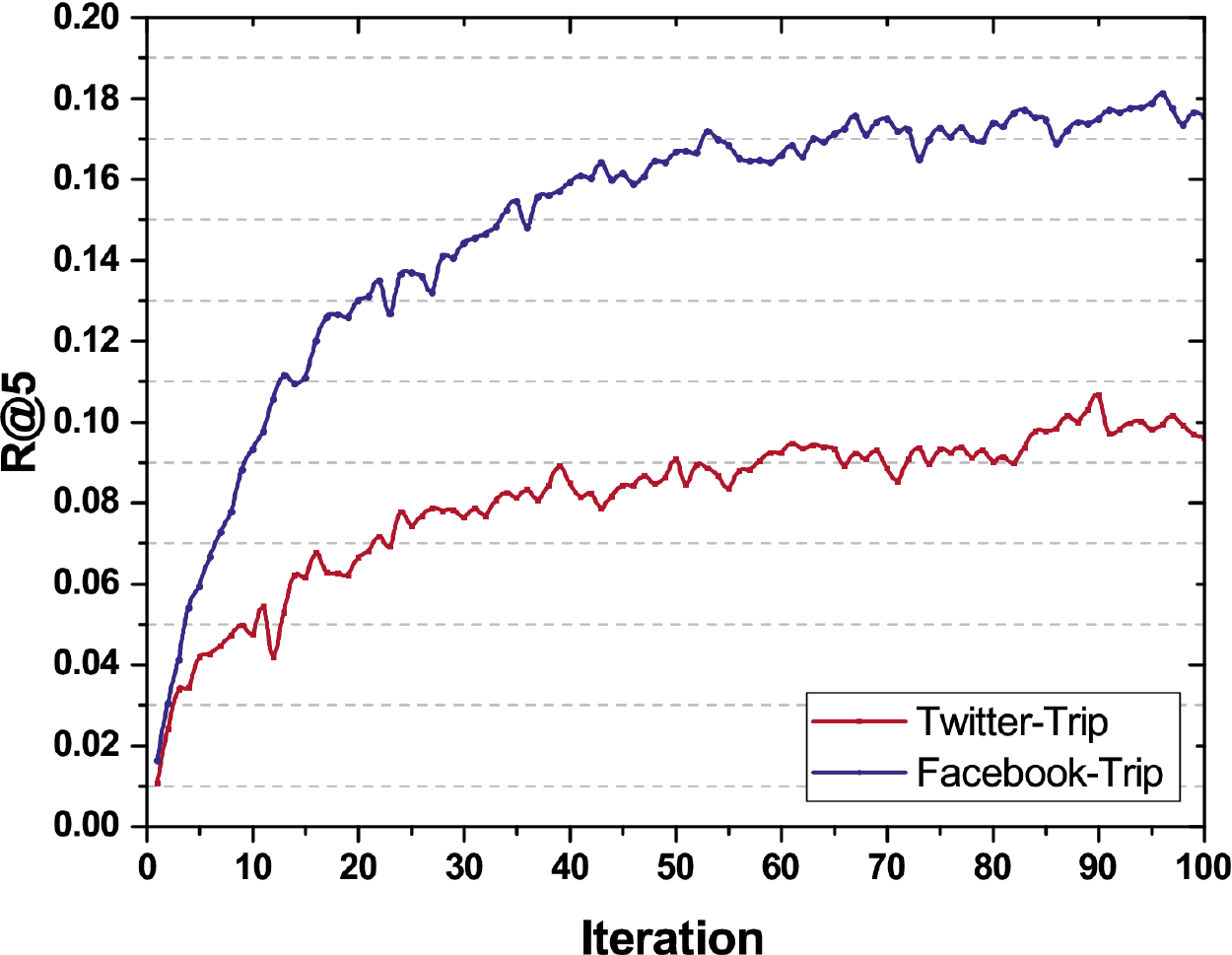}
			\label{fig:r-sensitivity}}
		\vspace{-10pt}
		\caption{Training loss and recommendation performance regarding AUC and R@$5$ \wrt the number of iterations.}
		\vspace{-1em}
		\label{fig:sensitivity-analysis}
	\end{figure*}
	
	\begin{figure*}[h]
		\centering
		\subfigure[AUC vs. dropout ratio $\rho$]{
			\includegraphics[width=0.235\textwidth]{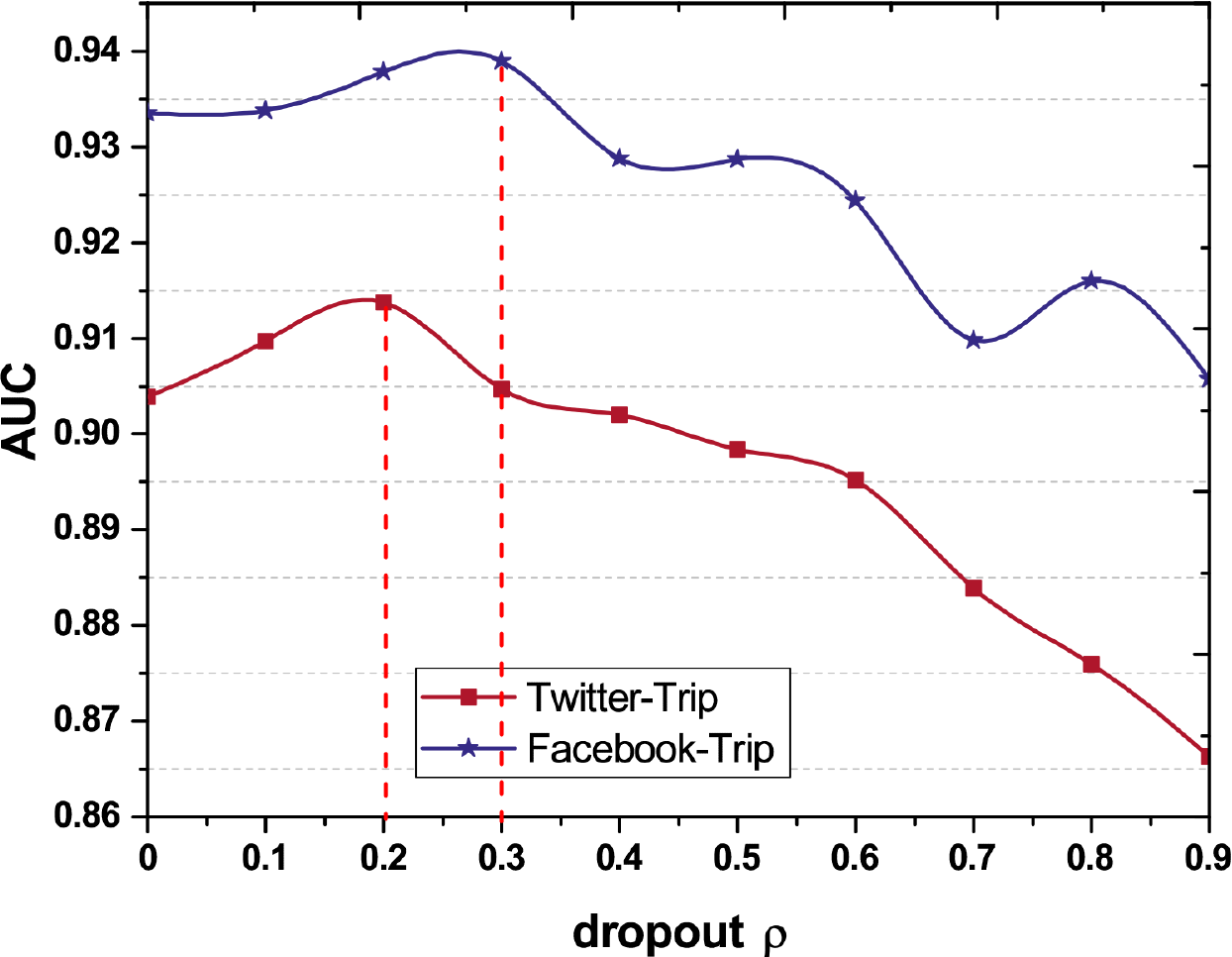}
			\label{fig:dropout-auc}}
		\subfigure[R@$5$ vs. dropout ratio $\rho$]{
			\includegraphics[width=0.235\textwidth]{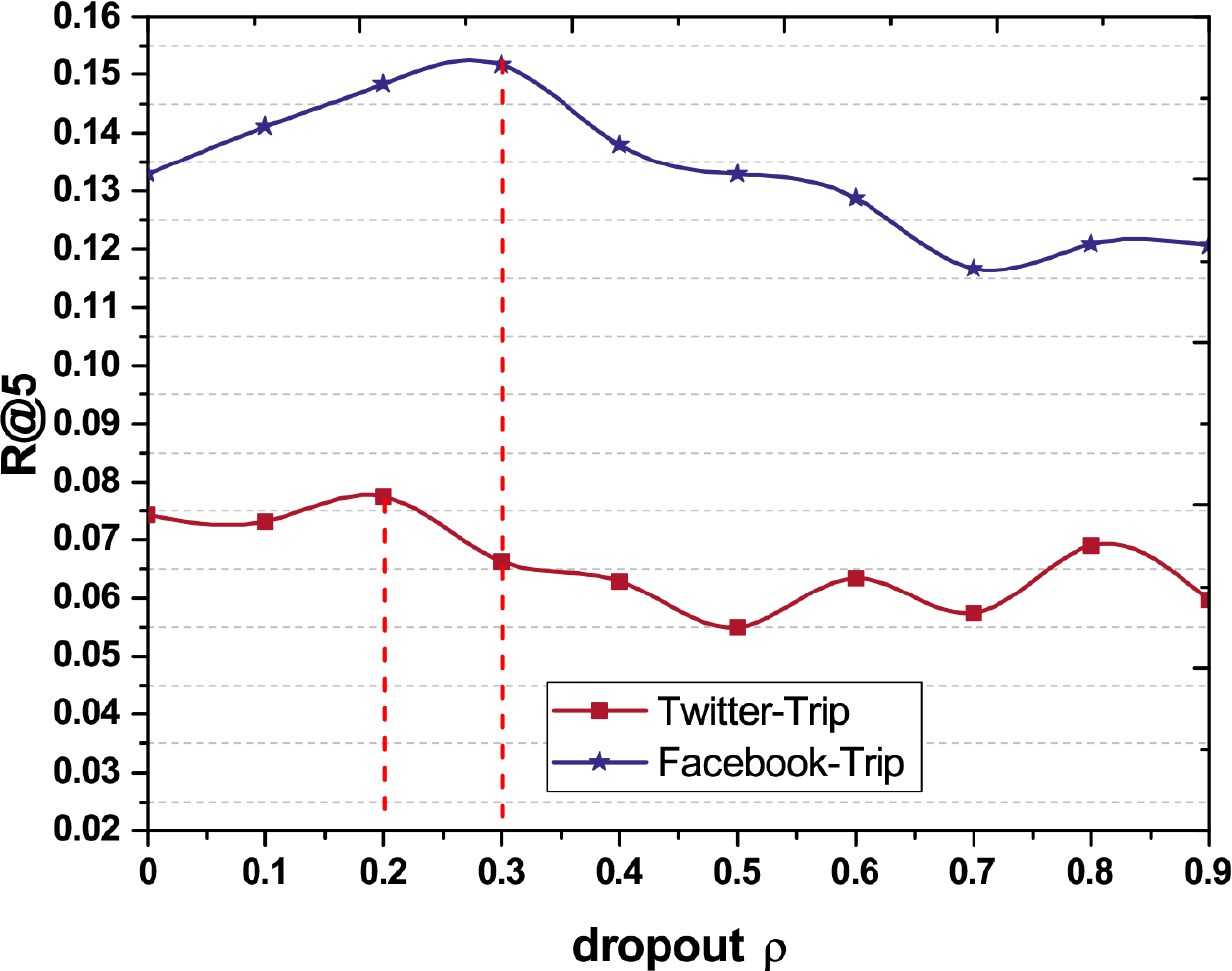}
			\label{fig:dropout-r}}
		\subfigure[AUC vs. tradeoff parameter $\mu$]{
			\includegraphics[width=0.235\textwidth]{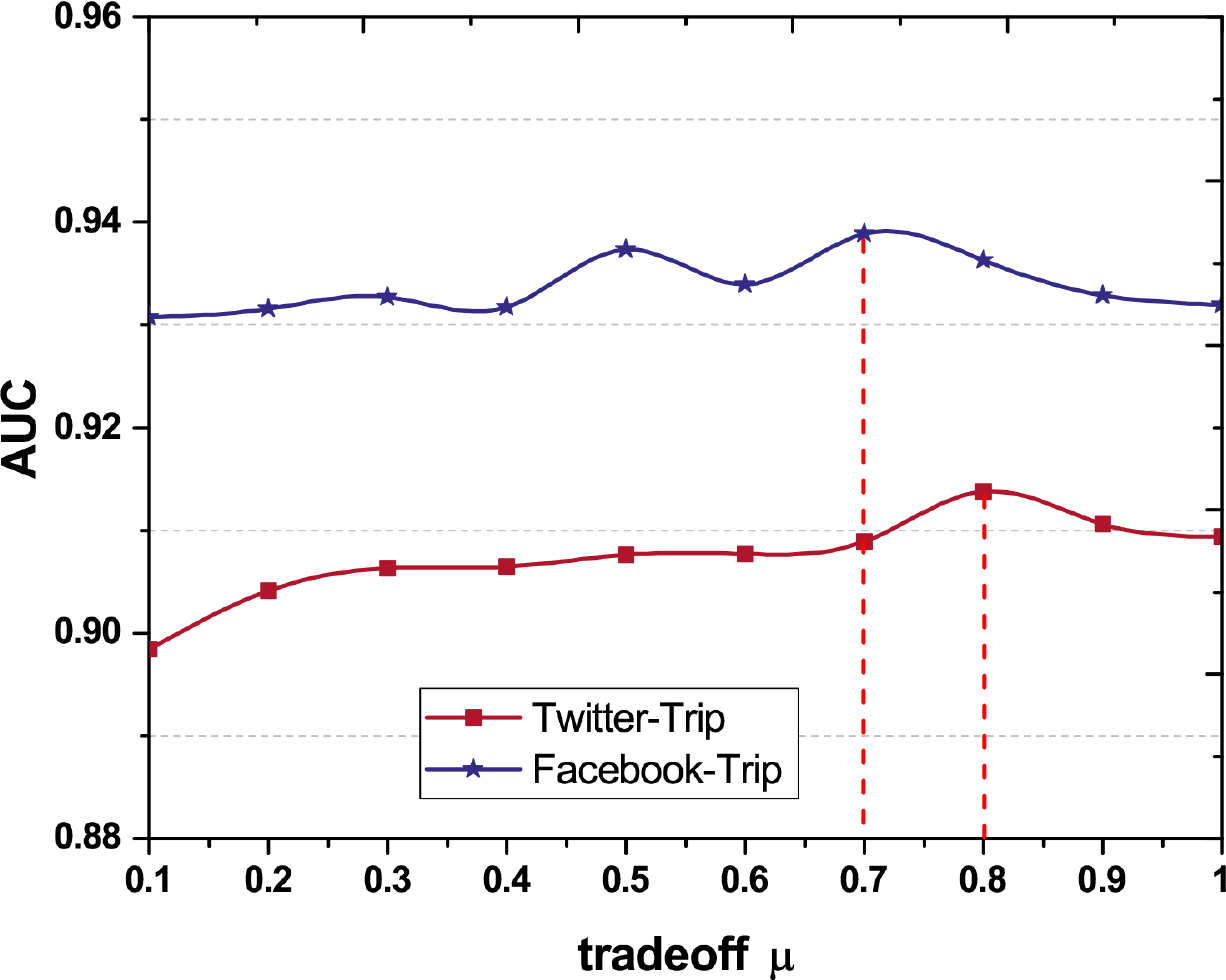}
			\label{fig:tradeoff-auc}}
		\subfigure[R@$5$ vs. tradeoff parameter $\mu$]{
			\includegraphics[width=0.235\textwidth]{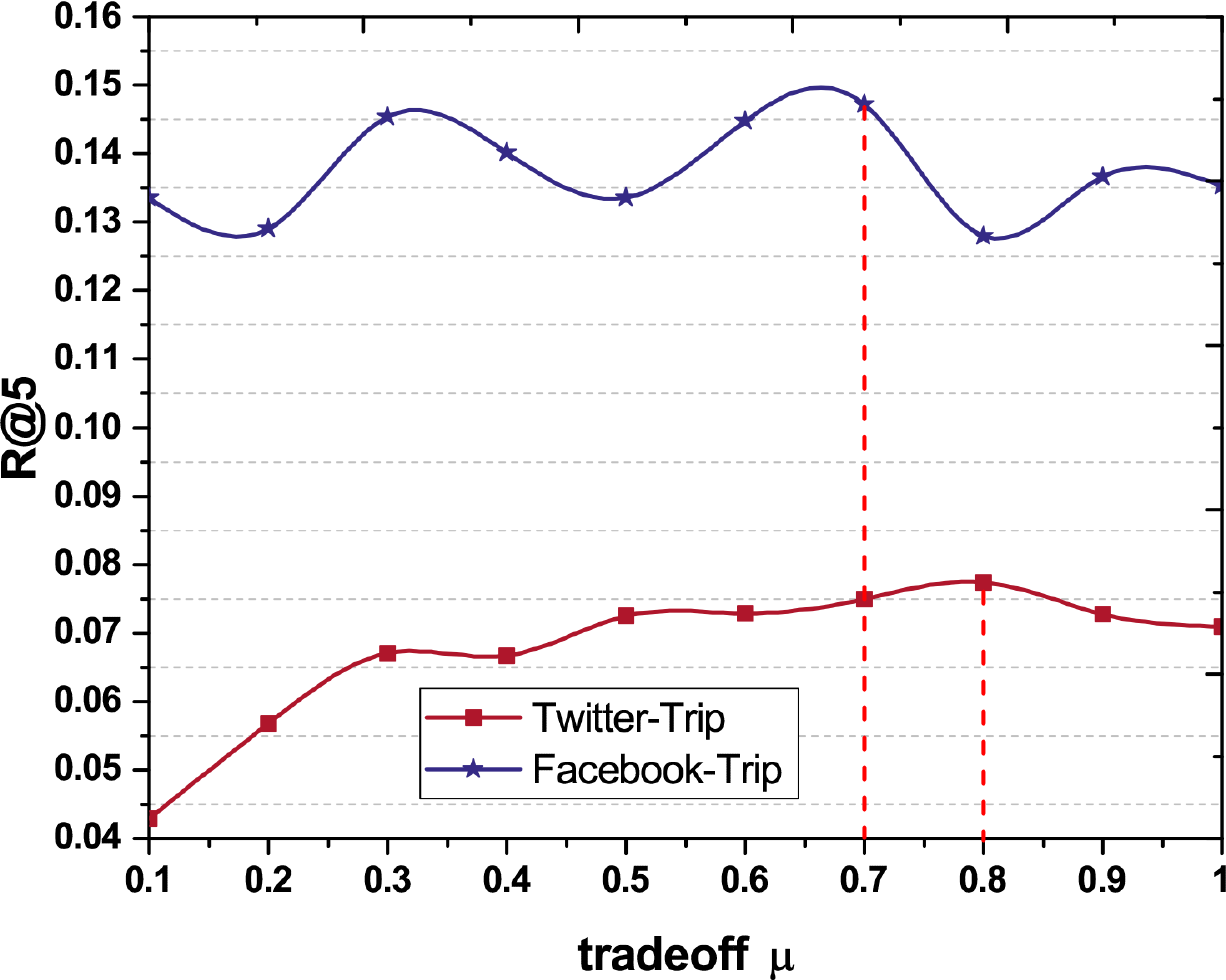}
			\label{fig:tradeoff-r}}
		\vspace{-10pt}
		\caption{Performance comparison of AUC and R@$5$ \wrt the dropout ratio $\rho$ and tradeoff parameter $\mu$ on Twitter-Trip and Facebook-Trip datasets.}
		\vspace{-1em}
		\label{fig:dropout-tradeoff}
	\end{figure*}
	
	\subsection{Study of NSCR (RQ2)}
	In this subsection, we empirically study the convergence of NSCR and then purpose to analyse the influences of several factors, such as dropout ratio and tradeoff parameter, on our framework.
	
	\textbf{Convergence:}~We separately present the training loss and the performance w.r.t. AUC and R@$5$ of each iteration in Figures~\ref{fig:training-loss},~\ref{fig:auc-sensitivity}, and~\ref{fig:r-sensitivity}. Jointly observing these Figures, we can see that training loss of NSCR gradually decreases with more iterations, whereas the performance is generally improved. This indicates the rationality of our learning framework. Moreover, the most effective updates occurs in the first $20$ iterations, which indicates that effectiveness of our learning framework. As Figure~\ref{fig:r-sensitivity} shows, the performance regarding R@$5$ fluctuates markedly over the iteration times, while that regarding AUC is quite stable. It is reasonable since R@$5$ only considers the top-$5$ results rather than the relative order as AUC defined.
	
	\textbf{Impact of Dropout:}~We employ the dropout technique in NSCR to prevent our model from overfitting, instead of regularizing model parameters. Figures~\ref{fig:dropout-auc} and~\ref{fig:dropout-r} present the performance \wrt AUC and R@$5$ of NSCR-$0$ by varying the dropout ratio $\rho$ on the pairwise pooling layer, respectively. As we can see, when dropout ratio equals to $0$, NSCR-$0$ suffers severely from overfitting. Moreover, using a dropout ratio of $0.3$ and $0.2$ leads to the best performance on Twitter-Trip and Facebook-Trip datasets, respectively. However, when the optimal dropout ratio exceeds the optimal settings, the performance of NSCR-$0$ greatly decreases, which suffers from insufficient information. This highlights the significance of using dropout, which can be seen as ensembling multiple sub-models~\cite{DBLP:journals/jmlr/SrivastavaHKSS14}.
	
	\textbf{Impact of Tradeoff Parameter:}~There is one positive parameter $\mu$ in the social modelling, which can capture the tradeoff between the fitting regularizer and the normalized graph Laplacian, as Eqn.\eqref{equ:social-training} shows. Figures~\ref{fig:tradeoff-auc} and~\ref{fig:tradeoff-r} present the performance \wrt. AUC and R@$5$, respectively. As we can see, setting $\mu$ of $0.8$ and $0.7$ can lead to the optimal performance on Twitter-Trip and Facebook-Trip datasets, respectively. And the performance of NSCR-$0$ changes within small ranges nearby the optimal settings. It justifies that our model is relatively insensitive to the parameter around its optimal configuration.
	
	\begin{table}[t]
		\centering
		\caption{Recommendation performance of NSCR with different hidden layers.}
		\vspace{-5pt}
		\label{tab:deep-performance}
		\resizebox{0.46\textwidth}{!}{\begin{tabular}{|c|c|c|c|c|c|c|}
				\hline
				\textbf{Metrics} & \multicolumn{3}{c|}{\textbf{AUC}}                         & \multicolumn{3}{c|}{\textbf{R@$\Mat{5}$}}                       \\ \hline
				\textbf{Factors} & \textbf{NSCR-$\Mat{0}$} & \textbf{NSCR-$\Mat{1}$} & \textbf{NSCR-$\Mat{2}$} & \textbf{NSCR-$\Mat{0}$} & \textbf{NSCR-$\Mat{1}$} & \textbf{NSCR-$\Mat{2}$} \\ \hline\hline
				\multicolumn{7}{|c|}{\textbf{Twitter-Trip}}                                                                                              \\ \hline
				\textbf{8}       & $0.8598$          & $0.8630$          & $\Mat{0.8704}$    & $0.0585$          & $0.0604$          & $\Mat{0.0628}$    \\ \hline
				\textbf{16}      & $0.8883$          & $0.8984$          & $\Mat{0.9026}$          & $0.0738$          & $0.0672$          & $\Mat{0.0812}$    \\ \hline
				\textbf{32}      & $0.9018$          & $0.9056$          & $\Mat{0.9109}$    & $0.0723$          & $0.0742$          & $\Mat{0.0843}$    \\ \hline
				\textbf{64}      & $0.9138$          & $0.9175$          & $\Mat{0.9222}$    & $0.0717$          & $0.0697$          & $\Mat{0.0725}$    \\ \hline
				\textbf{128}     & $0.9003$          & $0.9034$          & $\Mat{0.9125}$    & $0.0519$          & $0.0653$          & $\Mat{0.0688}$    \\ \hline\hline
				\multicolumn{7}{|c|}{\textbf{Facebook-Trip}}                                                                                             \\ \hline
				\textbf{8}       & $0.8978$          & $0.8922$          & $\Mat{0.9034}$    & $0.0860$          & $0.0872$          & $\Mat{0.0986}$    \\ \hline
				\textbf{16}      & $0.9165$          & $0.9197$          & $\Mat{0.9265}$    & $0.1048$          & $0.1388$          & $\Mat{0.1419}$    \\ \hline
				\textbf{32}      & $0.9303$          & $0.9322$          & $\Mat{0.9335}$    & $0.1441$          & $\Mat{0.1486}$    & $0.1465$          \\ \hline
				\textbf{64}      & $0.9337$          & $0.9376$          & $\Mat{0.9390}$    & $0.1353$          & $0.1359$          & $\Mat{0.1466}$    \\ \hline
				\textbf{128}     & $0.9270$          & $0.9310$          & $\Mat{0.9332}$    & $0.1168$          & $0.1304$          & $\Mat{0.1373}$    \\ \hline
			\end{tabular}}
			\vspace{-2em}
		\end{table}
		
		\subsection{Impact of Hidden Layer (RQ3)}
		To capture the complex and non-linear inherent structure of user-item interactions, we employ the a deep neural network for our task. It is curious whether NSCR can benefit from the deep architecture. Towards this end, we further investigate NSCR with different number of hidden layers. As it is computationally expensive to tune the dropout ratio $\rho$ for each hidden layer, we simply apply the same settings for all layers. The empirical results on two datasets are summarized in Table~\ref{tab:deep-performance} whereinto NSCR-2 indicates the NSCR method with two hidden layers (besides the embedding layer and prediction layer), and similar notations for others. We have the following observations:
		
		\begin{itemize}[leftmargin=*]
			\item In most cases, stacking more hidden layers is helpful for the recommendation performance. NSCR-$2$ and NSCR-$1$ achieve consistent improvement over NSCR-$0$, which has no hidden layers and directly projects the embedding to the prediction layer. We attributed the improvement to the high nonlinearity achieved by stacking more hidden layers. Our finding is consistent with \cite{DBLP:conf/cvpr/HeZRS16} and again verifies the deep neural networks have strong generalization ability. However, it is worth mentioning that such a deep architecture needs more time to optimize our framework and easily leads to the overfitting due to the limited training data in our datasets.
			\item Increasing the width of hidden layers (\ie the embedding size) from $8$ to $64$ can improve the performance significantly, as that of increasing their depth. However, with the embedding size of $128$, NSCR degrades the performance. It again verifies that using a large number of the embedding size has powerful representation ability~\cite{DBLP:conf/cvpr/HeZRS16}, but may adversely hurt the generalization of the model (\eg overfitting the data)~\cite{heneural,DBLP:conf/cvpr/HeZRS16}.
		\end{itemize}

\section{Related Work}
\subsection{Social Recommendation}
Social recommendation aims to leverage users' social connections to enhance a recommender system~\cite{DBLP:conf/wsdm/RenLLWR17,DBLP:conf/kdd/ZhangZYZHH16}.
It works by modelling social influence, which refers to the fact that a user's decision can be affected by her friends' opinions and behaviours.
%Existing studies work on incorporating social trust networks into recommender systems.
Ma \etal~\cite{DBLP:conf/wsdm/MaZLLK11} propose a social regularization term to enforce social constraints on traditional recommender systems. Based on a generative influence model, the work~\cite{DBLP:conf/sigir/YeLL12} exploits social influence from friends for item recommendation by leveraging information embedded in the user social network. The authors in~\cite{DBLP:conf/www/ZhangCYNL13} utilize social links as complementary data source to mine topic domains and employed domain-specific collaborative filtering to formulate users' interests. More recently,~\cite{DBLP:journals/tkde/JiangCCW0Y15} represents a star-structured hybrid graph centered at a user domain, which connects with other item domains, and transfers knowledge on social networks.

It is worth noting that the aforementioned studies are all based on social network relations of an information domain. While in this work, we focus on how to distill useful signal from an external social network~(\eg Facebook and Twitter), so as to improve the recommendation service of any information domain.

%within the user-item interactions itself; namely, they incorporate a social network of their own and do not rely on external social networks. It is more useful for real-world scenarios if we can glean useful data from an external social network, such as Facebook and Twitter.

\subsection{Cross-Domain Recommendation}
Distinct from the traditional recommendation methods that focus on data within a single domain, cross-domain recommendation concerns data from multiple domains. A common setting is leveraging the user-item interaction of a related auxiliary domain to improve the recommendation of the target domain. However, existing cross-domain recommendation work has an underlying assumption that the target and auxiliary domains are homogeneous. Depending on~\cite{DBLP:conf/www/ElkahkySH15,DBLP:journals/tkde/JiangCCW0Y15,farseev2017crossdomain}, they can be divided into two directions. One is assuming that different domains share overlapped user or item sets. The work~\cite{DBLP:conf/um/SahebiB13} augments ratings of movies and books for the shared users and accordingly conducts CF. Based on the shared users' latent space, the authors in~\cite{DBLP:conf/kdd/ChenHL13} leveraged cluster-level tensor sharing as a social regularization to bridge the domains. One more step, the authors in~\cite{DBLP:conf/www/HuCXCGZ13} formulated a generalized triadic user-item-domain relation over the common users and accordingly to capture domain-specific user factors and item factors.
%SPACE
%Sahebi \etal~\cite{DBLP:conf/recsys/SahebiB15} recommended items rated by the common users utilizing canonical correlation analysis across domains.
More recently, the authors~\cite{DBLP:conf/www/ElkahkySH15} proposed a multi-view deep learning recommendation system by using auxiliary rich features to represent users from different domains. Without aligned user or item, the other direction is on homogeneous data with the same rating scale. Codebook Transfer~\cite{DBLP:conf/ijcai/LiYX09} represents cluster-level rating patterns between two rating matrices in two related domains.
%SPACE
%Rating-matrix generative model~\cite{DBLP:conf/icml/LiYX09} extends the idea with a probabilistic model to solve collective transfer learning problems.
~\cite{DBLP:conf/kdd/TangWSS12} introduces a topic model to recommend authors to collaborate from different research fields.

Despite the compelling success achieved by previous work, little attention has been paid to recommendation across heterogeneous domains. In our settings, the source domain is a social network with user-user relations only, while the target domain is an information domain with user-item interactions. Hence, the auxiliary information is the social friendship, rather than the conventional interaction data. As a result, existing approaches can be hardly applied to this new research problem.

\section{Conclusion}

In this work, we systematically investigated cross-domain social recommendation, a practical task that has rarely been studied previously.
Towards this end, we proposed a generic neural social collaborative ranking (NSCR) solution, which seamlessly integrates user-item interactions of the information domain and user-user social relations of the social domain. To validate our solution, we constructed two real-world benchmarks of the travel domain, performing extensive experiments to demonstrate the effectiveness and rationality of our NSCR solution. The key finding of the work is that social signals contain useful cues about users' preference, even if the social signals are from social networks in a different domain. We achieved the goal by leveraging bridge users to unify the relevance signals from the two heterogeneous domains.
%We empirically verify that the social signal from bridge users are pivotal for our task, which can propagate the knowledge on users' preferences from the information domain and guide the item recommendation.
%Furthermore, we move an initial step towards leveraging deep neural networks to achieve our task. As such, we are capable of utilizing the rich attributed to enrich the user/item representations and capturing the complex and non-linear structure of user-item interactions.

Due to our restricted resources in collecting cross-domain data, the result is preliminary. Here we discuss several limitations of the current work, and our plans to address them in future.
First, in this work, we studied the recommendation performance of a travel-based information domain only, which is mainly for the ease of accessing the users' account on Facebook/Twitter. This results in a relatively small number of bridge users of our cross-domain datasets.
As a future work, we will collect a larger-scale set of data from the more popular information domains, such as E-commence sites, to explore the generalization ability of our solution to other information domains.
Second, due to the small number of bridge users, we forwent the study of user cold-start problem, as further holding out bridge users to simulate the cold-start scenario will pose challenge to the stability of evaluation.
With a larger-scale cross-domain data, we will study the effectiveness of our solution for cold-start users, as well as the influence of the bridge users' percentage.
Moreover, we restricted the SNSs by emphasizing only the social connections and omitting the weak user-item interactions in user-generated-contents. We will consider the weak user-item interaction in both domains to improve the recommendation performance.

%will pose stability challenge to the evaluation.

%we focus on only one travel information domain, which leads to several limitations.
%First, the bridge users are extremely scarce, which greatly limits the information propagation across domains. Furthermore, the current work ignores the cold-start problem since the scarce bridge users lead to our evaluation difficulty. In the future, we will collect large-scale data from multiple information domains, such as Ecommerce sites, and explore the generalization ability of our framework. Relying on the sufficient bridge users, in the future, we plan to focus on the cold-start problem, where the non-bridges can be treated as the cold-start users in the information domains, and further verify the effectiveness of our framework.

\noindent {\textbf{Acknowledgement}}
\noindent We would like to thank the anonymous reviewers for their valuable comments. NExT research is supported by the National Research Foundation, Prime Minister's Office, Singapore under its IRC@SG Funding Initiative.

\bibliographystyle{abbrv}
\balance

\bibliography{my-sigir}

\begin{thebibliography}{10}

\bibitem{iCD}
I.~Bayer, X.~He, B.~Kanagal, and S.~Rendle.
\newblock A generic coordinate descent framework for learning from implicit
  feedback.
\newblock In {\em WWW}, pages 1341--1350, 2017.

\bibitem{chen2017acf}
J.~Chen, H.~Zhang, X.~He, L.~N. and  Wei~Liu, and T.~Chua.
\newblock Attentive collaborative filtering: Multimedia recommendation with
  item- and component-level attention.
\newblock In {\em {SIGIR}}, 2017.

\bibitem{DBLP:conf/kdd/ChenHL13}
W.~Chen, W.~Hsu, and M.~Lee.
\newblock Making recommendations from multiple domains.
\newblock In {\em SIGKDD}, pages 892--900, 2013.

\bibitem{DBLP:conf/recsys/CovingtonAS16}
P.~Covington, J.~Adams, and E.~Sargin.
\newblock Deep neural networks for youtube recommendations.
\newblock In {\em {RecSys}}, pages 191--198, 2016.

\bibitem{DBLP:conf/www/ElkahkySH15}
A.~M. Elkahky, Y.~Song, and X.~He.
\newblock A multi-view deep learning approach for cross domain user modeling in
  recommendation systems.
\newblock In {\em WWW}, pages 278--288, 2015.

\bibitem{farseev2017crossdomain}
A.~Farseev, I.~Samborskii, A.~Filchenkov, and T.-S. Chua.
\newblock Cross-domain recommendation via clustering on multi-layer graphs.
\newblock In {\em {SIGIR}}, 2017.

\bibitem{fuli2017computational}
F.~Feng, L.~Nie, X.~Wang, R.~Hong, and T.-S. Chua.
\newblock Computational social indicators: a case study of chinese university
  ranking.
\newblock In {\em {SIGIR}}, 2017.

\bibitem{DBLP:conf/cvpr/HeZRS16}
K.~He, X.~Zhang, S.~Ren, and J.~Sun.
\newblock Deep residual learning for image recognition.
\newblock In {\em {CVPR}}, pages 770--778, 2016.

\bibitem{DBLP:conf/cikm/HeCKC15}
X.~He, T.~Chen, M.~Kan, and X.~Chen.
\newblock Trirank: Review-aware explainable recommendation by modeling aspects.
\newblock In {\em {CIKM}}, pages 1661--1670, 2015.

\bibitem{he2017neural}
X.~He and T.-S. Chua.
\newblock Neural factorization machines for sparse predictive analytics.
\newblock 2017.

\bibitem{heneural}
X.~He, L.~Liao, H.~Zhang, L.~Nie, X.~Hu, and T.-S. Chua.
\newblock Neural collaborative filtering.
\newblock In {\em WWW}, pages 173--182, 2016.

\bibitem{DBLP:conf/www/HuCXCGZ13}
L.~Hu, J.~Cao, G.~Xu, L.~Cao, Z.~Gu, and C.~Zhu.
\newblock Personalized recommendation via cross-domain triadic factorization.
\newblock In {\em WWW}, pages 595--606, 2013.

\bibitem{DBLP:journals/tkde/JiangCCW0Y15}
M.~Jiang, P.~Cui, X.~Chen, F.~Wang, W.~Zhu, and S.~Yang.
\newblock Social recommendation with cross-domain transferable knowledge.
\newblock {\em TKDE}, 27(11):3084--3097, 2015.

\bibitem{DBLP:conf/ijcai/LiYX09}
B.~Li, Q.~Yang, and X.~Xue.
\newblock Can movies and books collaborate? cross-domain collaborative
  filtering for sparsity reduction.
\newblock In {\em IJCAI}, pages 2052--2057, 2009.

\bibitem{SNE}
L.~Liao, X.~He, H.~Zhang, and T.-S. Chua.
\newblock Attributed social network embedding.
\newblock {\em arXiv preprint arXiv:1705.04969}, 2017.

\bibitem{DBLP:conf/wsdm/MaZLLK11}
H.~Ma, D.~Zhou, C.~Liu, M.~R. Lyu, and I.~King.
\newblock Recommender systems with social regularization.
\newblock In {\em WSDM}, pages 287--296, 2011.

\bibitem{DBLP:series/synthesis/2016NieSC}
L.~Nie, X.~Song, and T.~Chua.
\newblock {\em Learning from Multiple Social Networks}.
\newblock Synthesis Lectures on Information Concepts, Retrieval, and Services.
  Morgan {\&} Claypool Publishers, 2016.

\bibitem{DBLP:conf/wsdm/RenLLWR17}
Z.~Ren, S.~Liang, P.~Li, S.~Wang, and M.~de~Rijke.
\newblock Social collaborative viewpoint regression with explainable
  recommendations.
\newblock In {\em {WSDM}}, pages 485--494, 2017.

\bibitem{DBLP:conf/icdm/Rendle10}
S.~Rendle.
\newblock Factorization machines.
\newblock In {\em ICDM}, pages 995--1000, 2010.

\bibitem{DBLP:journals/tist/Rendle12}
S.~Rendle.
\newblock Factorization machines with libfm.
\newblock {\em TIST}, 3(3):57:1--57:22, 2012.

\bibitem{DBLP:conf/uai/RendleFGS09}
S.~Rendle, C.~Freudenthaler, Z.~Gantner, and L.~Schmidt{-}Thieme.
\newblock {BPR:} bayesian personalized ranking from implicit feedback.
\newblock In {\em UAI}, pages 452--461, 2009.

\bibitem{DBLP:conf/um/SahebiB13}
S.~Sahebi and P.~Brusilovsky.
\newblock Cross-domain collaborative recommendation in a cold-start context:
  The impact of user profile size on the quality of recommendation.
\newblock In {\em UMAP}, pages 289--295, 2013.

\bibitem{DBLP:conf/nips/SocherCMN13}
R.~Socher, D.~Chen, C.~D. Manning, and A.~Y. Ng.
\newblock Reasoning with neural tensor networks for knowledge base completion.
\newblock In {\em {NIPS}}, pages 926--934, 2013.

\bibitem{DBLP:conf/sigir/SongNZAC15}
X.~Song, L.~Nie, L.~Zhang, M.~Akbari, and T.~Chua.
\newblock Multiple social network learning and its application in volunteerism
  tendency prediction.
\newblock In {\em {SIGIR}}, pages 213--222, 2015.

\bibitem{DBLP:journals/jmlr/SrivastavaHKSS14}
N.~Srivastava, G.~E. Hinton, A.~Krizhevsky, I.~Sutskever, and R.~Salakhutdinov.
\newblock Dropout: a simple way to prevent neural networks from overfitting.
\newblock {\em {JMLR}}, 15(1):1929--1958, 2014.

\bibitem{RankALS}
G.~Tak\'{a}cs and D.~Tikk.
\newblock Alternating least squares for personalized ranking.
\newblock In {\em RecSys}, pages 83--90, 2012.

\bibitem{DBLP:conf/kdd/TangWSS12}
J.~Tang, S.~Wu, J.~Sun, and H.~Su.
\newblock Cross-domain collaboration recommendation.
\newblock In {\em SIGKDD}, pages 1285--1293, 2012.

\bibitem{DBLP:journals/tkde/WangFHLW17}
M.~Wang, W.~Fu, S.~Hao, H.~Liu, and X.~Wu.
\newblock Learning on big graph: Label inference and regularization with anchor
  hierarchy.
\newblock {\em {TKDE}}, 29(5):1101--1114, 2017.

\bibitem{DBLP:journals/tkde/WangFHTW16}
M.~Wang, W.~Fu, S.~Hao, D.~Tao, and X.~Wu.
\newblock Scalable semi-supervised learning by efficient anchor graph
  regularization.
\newblock {\em {TKDE}}, 28(7):1864--1877, 2016.

\bibitem{wang2017unifying}
X.~Wang, L.~Nie, X.~Song, D.~Zhang, and T.-S. Chua.
\newblock Unifying virtual and physical worlds: Learning toward local and
  global consistency.
\newblock {\em {TOIS}}, 36(1):4, 2017.

\bibitem{DBLP:conf/sigir/YeLL12}
M.~Ye, X.~Liu, and W.~Lee.
\newblock Exploring social influence for recommendation: a generative model
  approach.
\newblock In {\em SIGIR}, pages 671--680, 2012.

\bibitem{DBLP:conf/kdd/ZhangZYZHH16}
C.~Zhang, K.~Zhang, Q.~Yuan, L.~Zhang, T.~Hanratty, and J.~Han.
\newblock Gmove: Group-level mobility modeling using geo-tagged social media.
\newblock In {\em {SIGKDD}}, pages 1305--1314, 2016.

\bibitem{DCF}
H.~Zhang, F.~Shen, W.~Liu, X.~He, H.~Luan, and T.-S. Chua.
\newblock Discrete collaborative filtering.
\newblock In {\em SIGIR}, pages 325--334, 2016.

\bibitem{DBLP:conf/mm/ZhangZYYGC13}
H.~Zhang, Z.~Zha, Y.~Yang, S.~Yan, Y.~Gao, and T.~Chua.
\newblock Attribute-augmented semantic hierarchy: towards bridging semantic gap
  and intention gap in image retrieval.
\newblock In {\em {MM}}, pages 33--42, 2013.

\bibitem{DBLP:conf/www/ZhangCYNL13}
X.~Zhang, J.~Cheng, T.~Yuan, B.~Niu, and H.~Lu.
\newblock Toprec: domain-specific recommendation through community topic mining
  in social network.
\newblock In {\em WWW}, pages 1501--1510, 2013.

\bibitem{zhao2016user}
Z.~Zhao, H.~Lu, D.~Cai, X.~He, and Y.~Zhuang.
\newblock User preference learning for online social recommendation.
\newblock {\em TKDE}, 28(9):2522--2534, 2016.

\end{thebibliography}
\balance

\end{document}